\documentclass[aps,prl,twocolumn]{revtex4-1}
\usepackage{graphicx}
\usepackage{dcolumn}
\usepackage{amsmath}
\usepackage{latexsym}
\usepackage{mathrsfs}
\usepackage[usenames,dvipsnames]{color}
\usepackage{hyperref}
\usepackage{color}

\usepackage[normalem]{ulem}
\newcommand{\be}{\begin{equation}} 
\newcommand{\ee}{\end{equation}} 
\newcommand{\bea}{\begin{eqnarray}} 
\newcommand{\eea}{\end{eqnarray}} 

\begin{document}

\title{Site-percolation transition  of run-and-tumble  particles}

\author{
Soumya K. Saha, Aikya Banerjee, and P. K. Mohanty}
\email{pkmohanty@iiserkol.ac.in}
\affiliation {Department of Physical Sciences, Indian Institute of Science Education and Research Kolkata, Mohanpur, 741246 India.}
\begin{abstract}
We study percolation transition of run and tumble particles (RTPs)  on  a two dimensional  square lattice. RTPs in these models run to the  nearest neighbour  along their internal orientation  with unit rate,  and  to  other nearest neighbours with  rates  $p$.  In addition,  they  tumble to change their internal orientation  with rate $\omega$.  We show that for   small  tumble  rates,   RTP-clusters created  by joining occupied  nearest neighbours irrespective of their orientation  form a phase separated state  when  the rate of positional diffusion $p$  crosses a threshold; with further increase of $p$ the clusters disintegrate and another  transition to  a mixed phase occurs.  The critical exponents of  this re-entrant site-percolation transition of RTPs vary  continuously along the  critical line in the $\omega$-$p$  plane,  but a scaling function remains invariant.  This function    is identical to the corresponding  universal scaling  function  of  percolation transition  observed in the Ising model. We also show that the critical exponents of the underlying  motility induced phase separation transition   are related to  corresponding  percolation-critical-exponents  by  constant   multiplicative factors  known from the correspondence of magnetic and percolation critical exponents of the Ising model.\end{abstract}\maketitle

\section*{Introduction}
Active systems consume energy from the environment to produce self-propelled motion \cite{annurev:/content/journals/10.1146/annurev-conmatphys-070909-104101, Cates_2012, RevModPhys.85.1143,DEMAGISTRIS201565, RevModPhys.88.045006, annurev:/content/journals/10.1146/annurev-conmatphys-031218-013516}   and lead to  nonequilibrium steady states that exhibit collective behavior at many different length scales \cite{annurev:/content/journals/10.1146/annurev-conmatphys-031218-013516,doi:10.1073/pnas.0711437105, Ward2008-mw,Cavagna2017, Beer2020, PhysRevLett.108.098102}.  A specific kind  of self-propelled motion  performed by    certain bacteria and algae 
\cite{berg2003ecoli, doi:10.1126/science.1172667} are   described by a run and tumble dynamics, where particles   are assumed to have  a sense of  direction;  they  run  persistently  along  their internal  orientation and  tumble to change  their  orientation \cite{PhysRevLett.100.218103, Cates_2013, annurev:/content/journals/10.1146/annurev-conmatphys-031214-014710}. A common phenomenon unique to active matter systems is motility-induced phase separation (MIPS)  where the system  transits from a  mixed to a phase separated state (PSS)  with  increased  motility.   It is  widely believed    that motile particles having only excluded volume repulsion \cite{PhysRevLett.100.218103, PhysRevLett.108.235702, C3SM52469H, PhysRevLett.110.055701, PhysRevLett.111.145702, doi:10.1142/S0129183114410046, Suma_2014, PhysRevE.89.062301,Wittkowski2014, PhysRevLett.120.268003} can undergo  MIPS  transition. A stable  PSS in the absence of any attractive interaction is surprising and  understanding this phenomenon  has been a center of attention for many  researchers   in  recent years. Active matter systems are  modeled  theoretically  using hydrodynamic   descriptions \cite{RevModPhys.85.1143, PhysRevLett.108.235702, Bialke_2013}, 
agent based models  \cite{Schweitzer_2019, Ziepke2022} and lattice models  \cite{Thompson_2011, PhysRevLett.116.218101, Mallmin_2019, PhysRevE.102.062111,10.21468/SciPostPhys.14.6.165}. In experiments too, synthetically prepared self-propelled particles  \cite{RevModPhys.88.045006, PhysRevLett.104.138302,doi:10.1098/rsta.2013.0372, PhysRevE.108.034603, D4SM00305E} are  found to  exhibit collective motion.
 
In this article, we focus on  lattice models of RTPs. In one dimension (1D), RTPs  with a constant tumble rate  can not phase separate     \cite{10.21468/SciPostPhys.14.6.165}.  Models in 2D \cite{PhysRevE.89.012706, 10.1063/1.5023403, PhysRevE.92.042119, PhysRevE.94.022603, PhysRevE.98.030601} do exhibit phase separation  transitions. 
Characterization  of  critical behaviour  and universality class  of MIPS  transition is limited, although recent studies\cite{PhysRevLett.123.068002, D0SM02162H, Dittrich2021, Ray_2024} have claimed the transition to be in Ising universality class. Here, we take a different approach. We study site-percolation properties of RTP-clusters  on a square lattice,  and deduce  the  critical  behaviour  of the underlying MIPS phase transition.

\begin{figure}[h]
\centering
  \includegraphics[height=7cm]{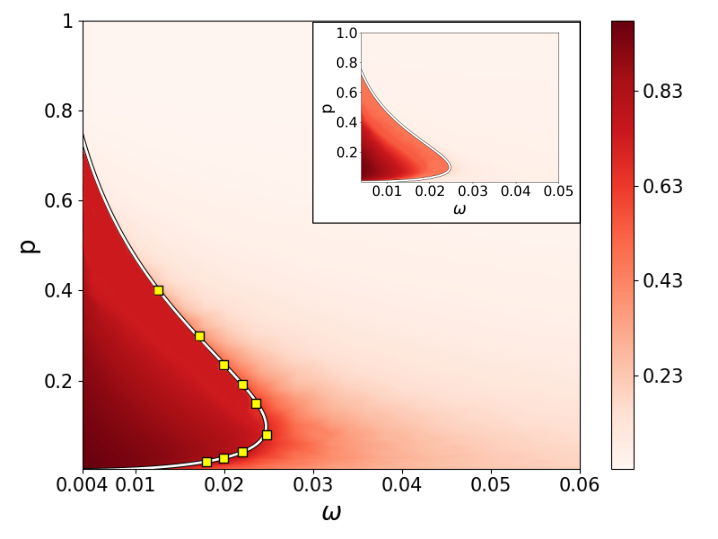}
  \caption{Density plot of  fraction of RTPs in largest cluster. The critical points (symbols) from Table \ref{table} are plotted along with the best fit critical line $p_c(\omega)$. The figure in the inset shows the density plot of the usual order parameter, defined in Eq. \eqref{eq:OP} of MIPS along with $p_c(\omega)$.}
  \label{fig:phases}
\end{figure}

During a percolation transition  at least one cluster starts becoming macroscopic in size  at  the critical point.  In  a particle conserved system,  formation of such a macro-cluster  
is bound to  create   a low density  region elsewhere. Thus  a phase separation is  expected  along with   the  percolation transition. However, the critical exponents  of  percolation  could be different  from that  of the MIPS  transition. 
In fact, in the context of  equilibrium phase transitions,  the site-percolation transition of the Ising  model occurs  exactly  at the same critical temperature where magnetic transition occurs, but their critical exponents differ \cite{PhysRevB.66.054107,PhysRevLett.62.1067}.  In 2D, the Ising critical exponents (for ferromagnetic transition)   is $\nu_{I}  =1,  \beta_{I}  = \frac18, \gamma_I= \frac74,$  whereas  critical exponents of the percolation transition \cite{PhysRevB.66.054107,PhysRevLett.62.1067} (characterized by average size of the  largest cluster) is $\nu_{P}  =1,  \beta_{P}  = \frac5{96}, \gamma_P= \frac{91}{48};$ 
the exponents  $\nu,\beta, \gamma$ are related to correlation length, order parameter and susceptibility respectively and subscripts $I,P$  stand  for Ising, Percolation.  Indeed percolation in Ising model form a  different universality class called  interacting percolation or  $Z_2$-percolation   ($Z_2$P) \cite{PhysRevB.66.054107,PhysRevLett.62.1067} which is different from the well known Ising universality class (IUC) in 2D. The ratio of  corresponding  exponents 
 \be \frac{\nu_{P}} {\nu_{I}}=1, ~ \frac{\beta_{P}}{ \beta_{I}} = 
 \frac{5}{12} 
 \label{eq:exp_pI}\ee 
 are  also universal  constants associated  with  the universality class \cite{PhysRevLett.62.1067}.

 In the RTP model we study in this article, the largest-cluster  exhibits a re-entrant percolation transition for small $\omega$  as we vary $p,$ the rate of  positional diffusion.  A macro-cluster  appears  when  $p$ is 
increased beyond a threshold, which disappears upon further increase of $p.$   The density plot of the  fraction of particles in the largest cluster is shown in Fig. \ref{fig:phases}. A representative critical line $p_c(\omega)$ passing through the critical points obtained from  numerical  simulations separates the two phases. The density plot of  usual order-parameter of MIPS  transition is shown in the  
inset  of   Fig. \ref{fig:phases} along with the  line $p_c(\omega)$  which appears to   differentiate  naturally the  PSS from the mixed one.    We find   that  the  critical  exponents  of the percolation  transition vary continuously along the  critical line (see Table \ref{table}) while a  scaling function remains invariant (see Fig. \ref{fig:xi2}) and  matched with the universal scaling function of $Z_2$P. Such a scenario is formally  termed as super-universality \cite{PhysRevLett.123.232002,PhysRevB.108.174417}. Thus, the site-percolation critical behaviour of  RTPs  form a  super-universality  class of  $Z_2$-percolation. We also find that the   critical  exponents of the underlying MIPS-phase transition  are related to  respective exponents of percolation  through  Eq. \ref{eq:exp_pI} and  form a superuniversality class of Ising model.  

\section*{The Model}
\begin{figure}[t]
\centering
\includegraphics[width=4.245cm]{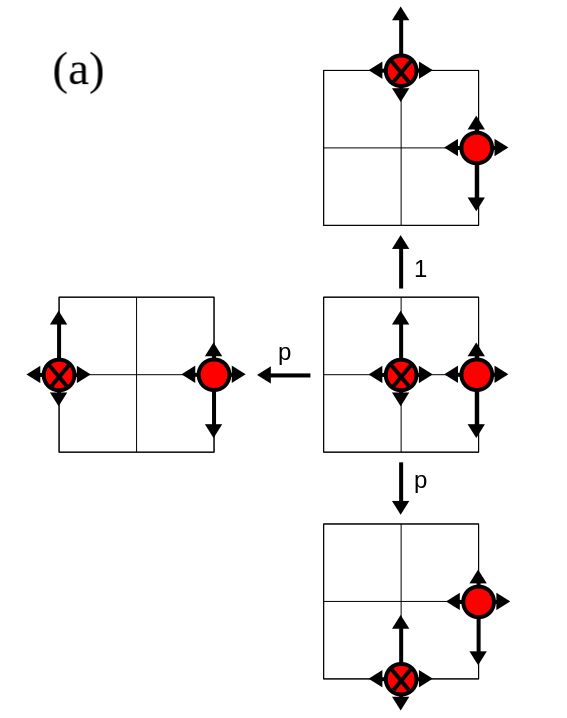}
\includegraphics[width=4cm]{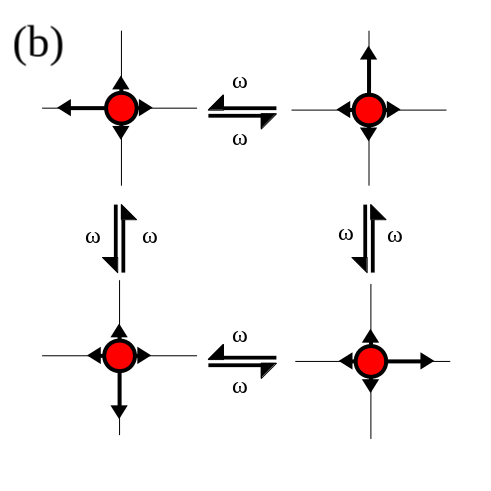}
\caption{ Dynamics of  RTPs  on a square lattice following Eq. \eqref{eq:dynamics}.  (a) Run:  A particle $m$ (marked by $\times$) has  internal orientation $\theta_m= \frac\pi2$ (upward) moves   from  site ${\bf i}$ to   the nearest neighbour  ${\bf i +\delta}_0$  (upward) with  unit rate, to site ${\bf i +\delta}_1$ (left) and    ${\bf i +\delta}_2$  (downward) with rate $p$; it can  not move to  ${\bf i +\delta}_3$ (right) as  the site is occupied by another  particle.  (b) Tumble: the particles can change their internal orientation $\theta_m$    to  $\theta_m  + \frac\pi2$ (rotate clockwise),  or  to   $\theta_m  - \frac\pi2$ (anticlockwise) with rates $\omega$.}
\label{fig:dynamics}
\end{figure}

We consider $N$  run and tumble particles  on  a square lattice with periodic boundary conditions in both directions, where sites labeled by ${\bf i}\equiv (x,y)$ with $x,y=1,2 \dots,L$  carry an occupation  index   $n_{\bf i}=0,1$ representing vacant and occupied sites respectively.  Each site  can  be  occupied by at most one particle respecting hard-core or excluded  volume  repulsion, and thus     $\sum_{\bf i} n_{\bf i} = N.$   The  particles are labeled by $m=1,2,\dots, N$  and each one carry   an internal  orientation  
 $ {\bf \theta}_m = 0, \frac\pi2, \pi, 3\frac\pi2,$ which represents  a  unit vector   pointing to one of the four neighbouring sites.
  The RTPs  are allowed to move (run) to the neighbour along their internal orientation with unit rate,  and to other three directions with rates $p,$  when the target site  is empty. If  the target site is occupied, such a move  is abandoned.   Particle  movement for a RTP  with  internal orientation $\theta= \frac\pi2$  is  depicted in  Fig. \ref{fig:dynamics}.  Here, run towards the  rightward neighbouring site  is  prohibited as it is occupied  by another RTP.In addition   particles  also tumble  with rate $\omega$   to  change their internal orientation  from ${\bf \theta}_m$  to  $ \theta_m \pm \frac\pi 2,$ as shown  in  Fig. \ref{fig:dynamics}.

The dynamics of the model  is  simulated as follows. A particle $m$ is chosen randomly  and independently from the collection of $N.$  Say, its position  is ${\bf i}$  and  internal orientation $\theta_m.$ In the Monte-Carlo simulation,  for small $\Delta t,$ the  following  events are attempted with different probabilities and  then,  time $t$  is increased by $\Delta t/N.$
\bea
 prob.~~ \Delta t&:&   {\rm if}~ n_{ {\bf i} + {\bf \delta}_0} =0, ~{\rm set}~ (n_{\bf i}=0,n_{ {\bf i} + {\bf \delta}_0} =1)  \label{eq1}\cr
prob.~ p \Delta t &:&   {\rm if}~ n_{ {\bf i} + {\bf \delta}_1} =0,~ {\rm set}~  (n_{\bf i}=0, n_{{\bf i} + {\bf \delta}_1} =1) \label{eq2}\cr
prob.~ p \Delta t &:&  {\rm if}~ n_{ {\bf i} + {\bf \delta}_2} =0, ~{\rm set}~  (n_{\bf i}=0, n_{{\bf i} + {\bf \delta}_2} =1) \label{eq3}\cr
prob.~ p \Delta t &:&    {\rm if}~ n_{ {\bf i} + {\bf \delta}_3} =0,~ {\rm set}~ (n_{\bf i}=0, n_{\bf i +\delta_3} =1) \label{eq4}\cr
prob.~ \omega \Delta t &:&  \theta_m  \to  \theta_m +\pi/2 \label{eq5}\cr
prob.~ \omega \Delta t &:&  \theta_m  \to \theta_m -\pi/2 \label{eq6},
\label{eq:dynamics}
\eea
and thus the configuration remains unchanged with probability 
$ 1- (1+3p+2\omega) \Delta t.$
 Here ${\bf \delta}_k = (\cos(\theta_m+ k \frac\pi2),   \sin(\theta_m+ k \frac\pi2))$  with $k=0,1,2,3$  are   the unit vectors pointing to nearest neighbours of  site ${\bf i}.$

 The dynamics of  this RTP model  is controlled by three parameters:  particle density $\rho= \frac{N}{L^2}$, the rate of  positional diffusion ($0<p<1$) and tumble rate ($\omega>0$).  It turns out that the   critical density $\rho_c$ is around  $\frac12$  for  any  $p,\omega$   (see Appendix for details) and  thus,  we primarily  focus  at  density $\rho=\frac12.$  Earlier numerical simulations \cite{PhysRevE.89.012706, 10.1063/1.5023403, Ray_2024,PhysRevE.89.062301}  have suggested  that MIPS  transition  is  not possible in  absence of positional diffusion, i.e., when $p=0$,  because    RTPs cannot escape from micro-clusters.  Again, for $p=1,$ the model reduces to a system of non-interacting hardcore  particles  which   move in all four directions with the same rate and thus  the system   remains  homogeneously mixed for any $\omega$.  Absence of an ordered state  at $p=0$ and $p=1$ necessarily indicate that  a phase separation  transition expected  for small $\omega$  must be re-entrant in the sense that  a PSS must appear  as $p$ is increased  and it  disappears with further increase of $p,$   which is evident from  Fig. \ref{fig:phases}.  In fact that MIPS might give rise to a re-entrant phase diagram has been theoretically predicted earlier within effective equilibrium theories \cite{PhysRevResearch.2.023207}.

  \section*{The  Percolation Transition}
  
  Now we  study  the  properties of the RTP clusters in the steady state. 
  Any configuration of $N$ RTPs can be viewed as  collection of  $K$-clusters, indexed as $k=1,2,\dots,K$  each containing $s_k$ number of particles, so that  $\sum_{k=1}^K  s_k =N.$  
 The clusters  are formed similar to those  in site-percolation problem \cite{STAUFFER19791, PhysRevE.71.036703, Essam_1980} where two occupied nearest neighbours belong to the same cluster  irrespective of their  internal orientations.    

 In a mixed state,  RTPs  are expected to form  small clusters  whereas in the PSS, there  must be at least one  macro-size cluster containing a finite-fraction of  the total $N$ particles (supplementary material \textsuperscript{\dag}). In  an infinite lattice,  the marco-cluster  is  an infinite-cluster  that  spans the lattice.
Thus, like site-percolation  transition  of   Ising and Potts models \cite{PhysRevB.66.054107, PhysRevE.71.036703} which occurs  at the  same  critical temperature where 
ferro-magnetic transition occurs,  one expects  that  RTPs  undergo a  percolation transition  whenever a phase separation transition occurs.  The percolation transition  is usually  characterized \cite{STAUFFER19791, PhysRevB.66.054107, MARGOLINA198273} by   an  order parameter $\phi$ and its  variance, namely the  susceptibility $\chi:$
 \be
 \phi= \frac 1{N} \langle s_{max}\rangle; \chi =\frac{1}{N^2} \left(\langle s_{max}^2  \rangle -  \langle s_{max}\rangle^2\right),
 \ee
 where $\langle.\rangle$ denotes the steady state average and  
$s_{max}$  is  the  number of particles in the largest cluster.
\begin{figure}[t]
\centering
\includegraphics[width=4.245cm]{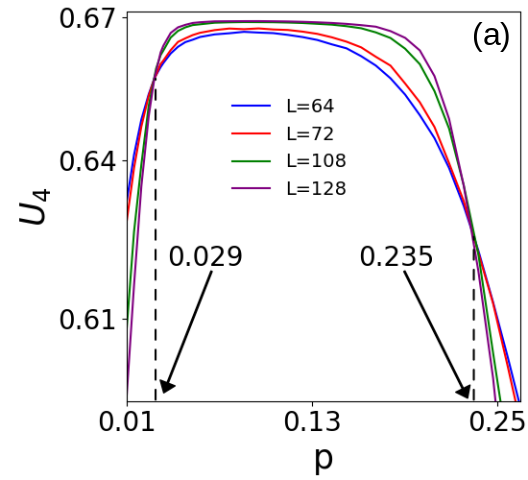}
\includegraphics[width=4.245cm]{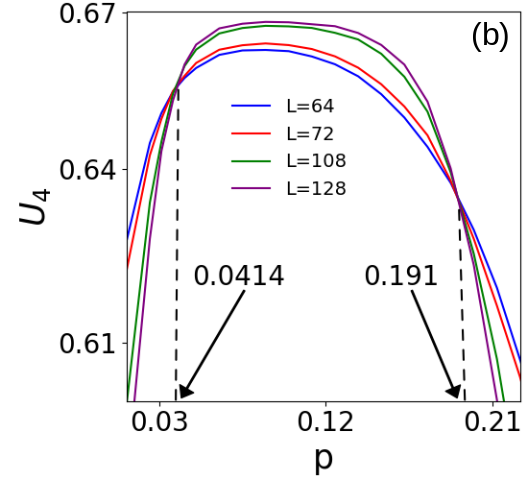}
\caption{(Color online) Binder Cumulant $U_4$ as a function of $p$ for (a) $\omega=0.020$ and (b) $\omega=0.022$. The two critical points for each $\omega$, marked with arrows indicate a re-entrant percolation transition. }
\label{fig:BC}
\end{figure}

\begin{figure}[t]
\centering
\includegraphics[width=4.25cm]{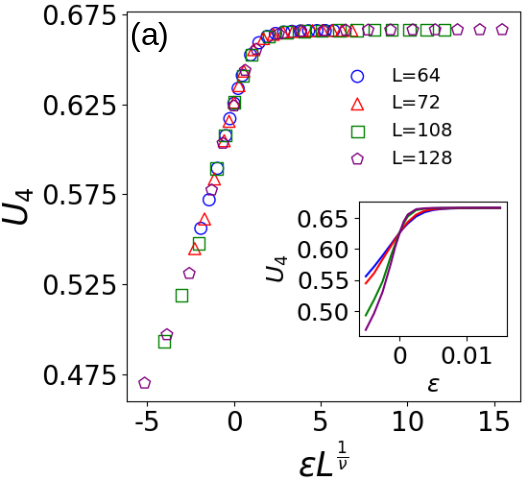}
\includegraphics[width=4.25cm]{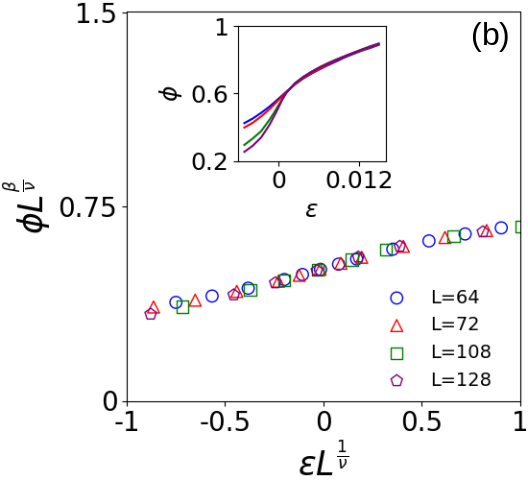}
\includegraphics[width=4.25cm]{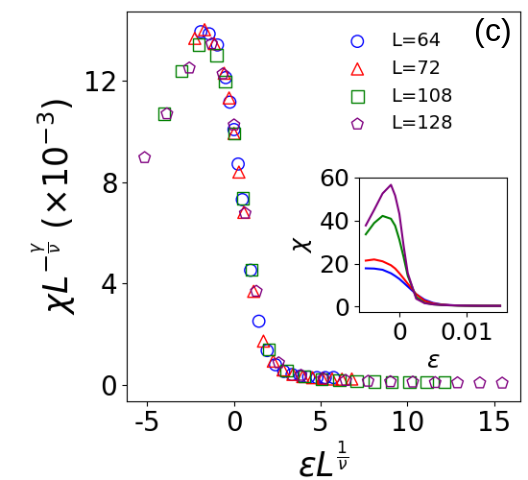}
\includegraphics[width=4.25cm]{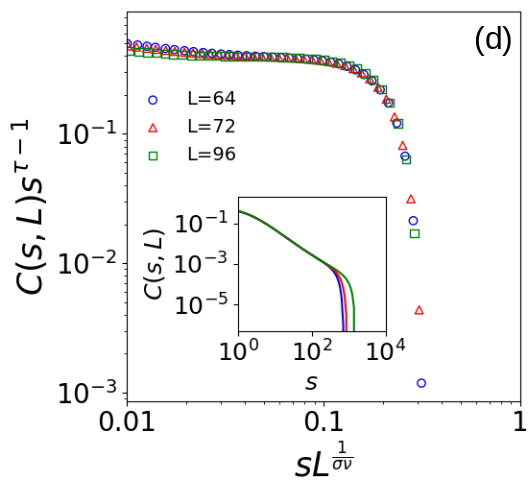}
\caption{(Color online) Finite size scaling at the critical point I $(p_c,\omega_c)=(0.235,0.020)$. (a) $U_4$, (b) $\phi L^\frac{\beta}{\nu}$ and (c) $\chi L^\frac{-\gamma}{\nu}$ vs. $\varepsilon L^{\frac{1}{\nu}}$ where $\varepsilon = \omega_c-\omega$. The best collapse are obtained for $\frac{1}{\nu}=1.43,\frac{\beta}{\nu}=0.142,\frac{\gamma}{\nu}=1.725$. (d) Scaling collapse of $C(s,L)s^{\tau-1}$ vs. $sL^\frac{1}{\sigma\nu}$ yields the exponents $\tau=2.075$ and $\frac{1}{\sigma\nu}=1.86$. Insets show the raw data. }
\label{fig:p0.235}
\end{figure}

From the Monte Carlo simulations of the system  at density 
$\rho=\frac{1}{2},$ we measure $\phi,$   $\chi$  and the  Binder-cumulant 
\begin{equation}
U_4=  1-  \frac{\langle s_{max}^4 \rangle }{3 \langle s_{max}^2 \rangle^2}
\label{eq:BC}
\end{equation} 
for different $p$ and $\omega.$  These calculations are repeated for different system sizes  and the critical exponents are determined    from  the finite size scaling analysis \cite{binder2010monte, PhysRevLett.47.693,doi:10.1142/1011} described below.

 Binder cumulant  is independent of the system size at the critical point \cite{Landau_Binder_2014, PhysRevLett.88.185701} 
and  thus, the   intersection  point of $U_4$ versus $p$ (or $\omega$) curves  for  different $L$  provide an estimates  of $p_c$ (or $\omega_c).$ Figure \ref{fig:BC} describes this for
$\omega=0.020, \omega= 0.022$.  In both cases,  $U_4$ versus $p$ curves  show two intersection points indicating  a re-entrant  transition. In Fig.  \ref{fig:BC}(a)  transition from  a mixed phase  to PSS occurs at $p_c=0.029$ and  PSS  to a mixed phase occurs  again at   at $p_c=0.235.$  For  $\omega=0.022,$  the transitions occur at  $p_c=0.0414$ and $0.191$  respectively. Other estimated  values of $(\omega_c, p_c)$ are listed in Table \ref{table}.

\begin{table}[!ht]
    \centering
    \caption{Critical points and exponents of percolation transition of RTPs in 2D}
    \label{table}
    \begin{tabular}{cccccc}
        \hline
        Sl. No. & $p_c$ & $\omega_c$ & $1/\nu$ & $\beta/\nu$ & $\gamma/\nu$ \\ \hline
        I & 0.235 & 0.020(0) & 1.43 & 0.14(2) & 1.72(5) \\
        II & 0.150 & 0.023(5) & 1.26 & 0.10(1) & 1.75(3) \\ 
        III & 0.080 & 0.024(7) & 1.22 & 0.09(2) & 1.82(4) \\
        IV & 0.0290 & 0.020(0) & 1.13 & 0.06(6) & 1.86(8) \\ 
        V & 0.0275 & 0.019(8) & 1.11 & 0.06(5) & 1.87(2) \\
        VI & 0.020 & 0.018(0) & 1.10 & 0.05(5) & 1.89(5) \\ \hline  
        $Z_2$P \cite{PhysRevLett.62.1067}& - & - & 1 & $\frac{5}{96}\simeq$ 0.052 & $\frac{91}{48} \simeq$ 1.896 \\ \hline
    \end{tabular}
\end{table}

\begin{figure}[h]
\centering
\includegraphics[width=8cm,height=6 cm]{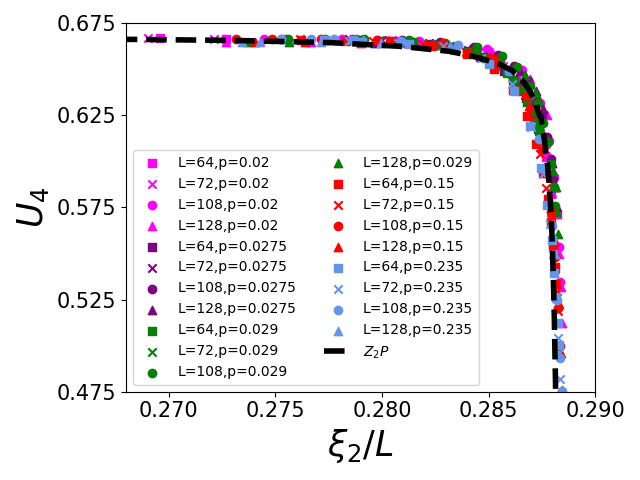}
\caption{ (Color online) Plot of $U_4$ vs. $\xi_2/L$ parameterized by $\omega$ for different $p$ and $L$. The dashed line corresponds to the same scaling function of $Z_2$P obtained from simulation of Ising percolation.}
\label{fig:xi2}
\end{figure}

Now we  vary  one of the parameters $(p,\omega)$  about the critical value  $(p_c,\omega_c)$ and  calculate  $\phi, \chi, U_4$  for different  $L$ using  Monte Carlo simulations. Using their  finite size scaling properties \cite{binder2010monte, PhysRevLett.47.693,doi:10.1142/1011},
 \begin{eqnarray}\label{eq:scaling_binder}
 \phi = L^{-\frac{\beta}{\nu}}f_\phi(\varepsilon L^{\frac1\nu}); ~ 
 \chi = L^{\frac\gamma\nu}f_\chi(\varepsilon L^{\frac1\nu});~
  U_4 = f_{b}(\varepsilon L^{\frac1\nu}),
\end{eqnarray}
where  $\varepsilon$  is a  measure of distance from the critical point and $f_{\phi, \chi,b}(.)$ are universal scaling functions,   we  obtain the exponent ratios   $\frac1\nu,$ $\frac\gamma\nu$ and  $\frac\beta\nu$ as  the fitting parameters that result in the best  scaling collapse. The estimated critical  exponents are listed in Table \ref{table}. For  demonstration, we  choose  $p=0.235$  and vary  $\omega= \omega_c+\varepsilon$   to calculate $\phi, \chi, U_4$ for different  $L.$ A plot of $U_4, \phi L^{\frac\beta\nu}$ and $\chi L^{-\frac\gamma\nu}$ as a function of  the dimensionless parameter $\varepsilon L^{\frac1\nu}$  are shown  in Fig. \ref{fig:p0.235}(a),(b),(c) respectively.  The value  of $\frac1\nu=1.43$  that resulted   in the best collapse of $U_4$  in  Fig. \ref{fig:p0.235}(a)  is used in   Figs. \ref{fig:p0.235}(b) and (c)  to obtain the best collapse  for $f_\phi(.),f_\chi(.)$   by tuning  $\frac\beta\nu$ and $\frac\gamma\nu$ respectively.   The estimated  values  are  $\frac\beta\nu=0.14(2)$ and  $\frac\gamma\nu=1.72(5).$ 

To study  the  cluster properties of RTPs we notice that in the near-critical  regime,  the distribution of finite clusters $P(s)$  follow a scaling relation $P(s,\varepsilon) =  s^{-\tau} f(\varepsilon s^\sigma)$ where  the exponents  $\tau,\sigma$ 
 obey  the  scaling relations \cite{STAUFFER19791}, 
\be \tau = 2 +\frac\beta{\beta+\gamma},~ \sigma^{-1} =\beta+\gamma, \label{eq:tau_sig} \ee
  The fractal dimension   of the largest cluster (which is the 
 spanning-cluster in the thermodynamic limit)  is $d_f =d- \frac{\beta}{\nu}$ \cite{Vanderzande_1992},  which can also be expressed  as $d_f = \frac{d}{\tau-1}$ using the scaling relation $2\beta+ \gamma = d \nu.$  

In finite systems, the correlation length is limited by $L$, resulting in $\varepsilon  \sim L^{-\frac1\nu}.$ Thus, the  probability  of finding clusters  of size $s$ or more is 
\be 
C(s, L ) \equiv  \sum_{s'=s}^\infty P(s',L)  = s^{1- \tau} g(sL^{\frac1{\sigma\nu}}).
\ee
From the Monte Carlo simulations we obtain $C(s,L)$  at  the critical point I $(p_c,\omega_c)=(0.235,0.020)$ for  different $L$ and plot   $C(s,L) s^{\tau-1}$ as a function of $sL^{\frac1{\sigma\nu}}$ in Fig. \ref{fig:p0.235}(d). We use   $\tau = 2.075$ and $\frac1{\sigma\nu}=1.86,$  calculated from Eq. \eqref{eq:tau_sig} and Table \ref{table}.
A  good collapse  observed here  assures   that   the critical exponents obey  the known  scaling relations of  percolation phenomena \cite{STAUFFER19791, Essam_1980}. The critical exponents for the other critical points II to VI in Table \ref{table} are calculated in a similar way, and described in ESI\textsuperscript{\dag}.

\subsection*{Universality}

Typically, the universality class of a phase transition is identified by a unique set of critical exponents. However, the critical exponents of the percolation transition of RTPs, as shown in Table \ref{table} , vary continuously along the critical line, adhering to the scaling relation $2\beta+\gamma= d\nu$ with $d=2,$ within the error limits.  Continuous variation  of critical exponents is not new   to the study of equilibrium phase transitions; it can  arise due to  presence of a marginal operator. In the context of the Ashkin-Teller model, where the critical exponents of both magnetization and polarization transitions vary continuously, it was observed  \cite{PhysRevLett.123.232002, PhysRevB.108.174417} that  a  RG-invariant scaling function   $U_4 = F\left(\frac{\xi_2}L\right)$ remains   invariant  along the critical line; this scaling function  relates   the Binder cumulant  $U_4$   with   second-moment correlation length $\xi_2,$ where

  \be (\xi_2)^2 =\frac{ \int_0^\infty  r^2  g(r)dr}{ \int_0^\infty g(r)dr};  ~~g(r) = \langle n_{\bf i} n_{\bf i+r}\rangle -  \langle n_{\bf i} \rangle \langle n_{\bf i+r}\rangle .\ee
Based on this,  it was  proposed that the  critical behaviour    of Ashkin-Teller model   belongs to the  Ising   superuniversality  class.

The   percolation   transition of RTPs   may also form a super-universality class - but which one ? We  notice that the  numerical value of the exponents  for small $(p,\omega)$  (say, critical point VI)  is  very  close to  the   exact  values    known for the  $Z_2$- percolation transition ($Z_2$P  in Table  \ref{table}) observed in Ising model \cite{PhysRevB.66.054107, PhysRevLett.62.1067}.  It  suggests   that  if  there  is a superuniversality  class,  it   must be   the  $Z_2$P. 
To verify this we  obtain $\xi_2$ and $U_4$  as  functions of $\omega$   using   Monte Carlo simulations, for different $p,L.$  The plots of    $U_4$  vs. $\frac{\xi_2}L$    for many different $(p,\omega, L)$  values plotted in Fig. \ref{fig:xi2}  falls on a universal function $F(.)$  which is no different from  the same   obtained  for  $Z_2$P universality class (dashed line).
We conclude that  the percolation  transition of  RTPs  belong   to the  super-universality class of  $Z_2$-percolation.

\subsection*{Connection to MIPS}
It is well known  \cite{PhysRevB.66.054107,PhysRevLett.62.1067} that the percolation transition of  Ising model carries the  signature  of the underlying  
magnetic  phase transition.  Specifically, both transitions occur at the same critical temperature, and their exponents are related as described by Eq. (\ref{eq:exp_pI}). Therefore, since the percolation transition of RTPs falls within the $Z_2$P universality class, it is likely that the underlying MIPS transition also occurs at the same critical point, with its critical exponents similarly related:  $ \beta' =  \frac{12}5\beta, \nu' =\nu$ and   $\gamma' =  d \nu' - 2\beta'$  (from hyper-scaling relations).

 \begin{figure}[h]
\centering
\includegraphics[width=4.25cm]{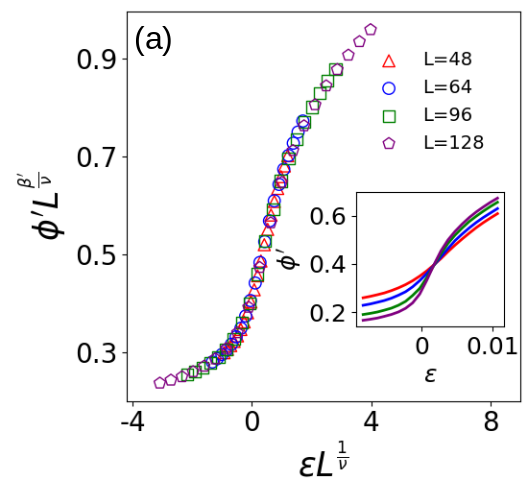}
\includegraphics[width=4.25cm]{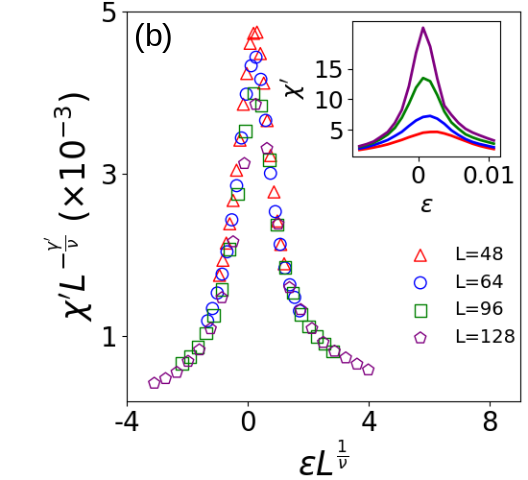}
\includegraphics[width=4.25cm]{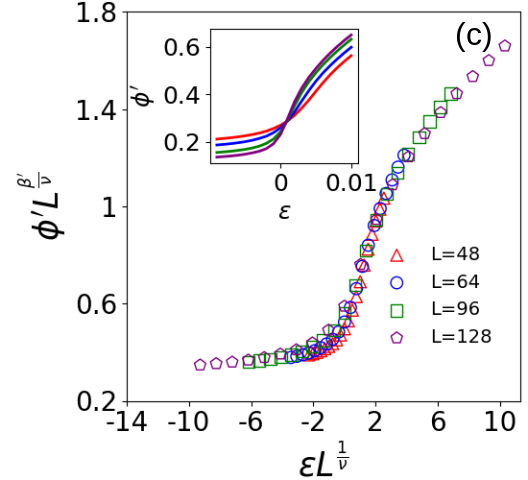}
\includegraphics[width=4.25cm]{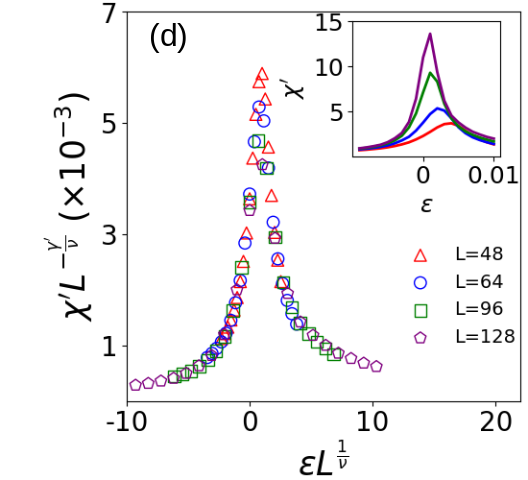}
\caption{(Color online) Finite size scaling collapse of MIPS order parameter $\phi'$ and susceptibility $\chi'$. (a), (b) correspond to the critical point III resulting in $\frac{\beta'}{\nu}=0.216,\frac{\gamma'}{\nu}=1.78.$ The same for the critical point I are shown in (c), (d) with resulting exponents $\frac{\beta'}{\nu}=0.336,\frac{\gamma'}{\nu}=1.66.$ }\label{fig:tphi}
\end{figure}

To verify, we  consider a rectangular system   with $L_x=L,  L_y=\frac L2$   and   study phase separation transitions  about the  critical points  I and III in Table \ref{table}  using an order parameter similar to one discussed in Refs. \cite{Ray_2024,PhysRevB.35.3372,PhysRevLett.88.145701},
\be 
 \phi' = \frac{2}{L_x L_y} \sum_{x=1}^{L_x} \left| N_{x} - \rho L_y\right|;~ N_{x}= \sum_{y=1}^{L_y} n_{x,y},
\label{eq:OP}
\ee
where $N_x$  is  the total number of particles at lattice sites ${\bf i}\equiv (x,y)$ with the same $x$-coordinate.  From Monte Carlo simulation of the model we calculate $\bar \phi =\langle  \phi' \rangle$ and $\bar \chi =\langle  \phi'^2\rangle -  \langle \phi'\rangle^2$  as a function of $\omega,$ keeping $p$ fixed.  The  plots in   Fig. \ref{fig:tphi}  shows  that  the data for different $L$ values  collapse   following  the finite size scaling,
 \begin{eqnarray}
 \phi' = L^{-\frac{ \beta'}{ \nu} }f'_\phi(\varepsilon L^{\frac1 \nu}); ~ 
 \chi' = L^{\frac {\gamma'}\nu} f'_\chi(\varepsilon L^{\frac1\nu}),
\end{eqnarray}
where we use $ \beta'=  \frac{12}{5} \beta,  \gamma' = 2\nu - 2 \beta'$  with    $ \frac1\nu,  \frac\beta\nu, \frac\gamma\nu$   taken from  Table \ref{table}. Note that critical points of the MIPS transition are taken to be the same as the corresponding percolation transition because it is evident from the density plot of  $\phi'$ in the inset of Fig. \ref{fig:phases} that the percolation critical line naturally separates the mixed state from the PSS. Thus,   the MIPS transition of RTPs in $\omega$-$p$ plane  lead to   continuous variation  of the  critical exponents  $ \beta',  \gamma',  \nu'$   obeying the scaling  relation $d\nu' = 2 \beta' + \gamma'.$ Thus, we believe that  the MIPS  transition  of RTPs in 2D belong to  Ising superuniversality class.  Note that  as $(\omega,p)\to (0,0)$  the  critical exponents  of MIPS transition approach  the Ising exponents following Eq. (\ref{eq:exp_pI}). This result is consistent with the Ising critical behavior reported in  RTP models studied earlier for small $p,\omega$  \cite{PhysRevLett.123.068002,D0SM02162H, Dittrich2021}.

\section*{Conclusion}
In conclusion, we study   percolation  transition of  run and tumble particles  on a square lattice, where  clusters  are formed by  joining occupied nearest neighbours  irrespective of their internal orientation. We find that   the system with a small tumbling rate $\omega$ undergoes a re-entrant percolation transition when the rate of   positional diffusion $p$ is increased. The transition belongs to the super-universality class of $Z_2$ percolation: all critical exponents of the transition vary continuously along the   critical line in $\omega$-$p$ plane but a scaling function $U_4=F\left(\frac{\xi_2}{L}\right)$ remains invariant. Thus the percolation transition belongs to the superuniversality class of $Z_2$P. Since any phase separated state  must contain  at least one macro cluster,  motility induced phase separation  transitions of RTPs   must  occur at the   critical point of percolation-transition, and it does. The critical exponents, however, differ and they are related to percolation  exponents by a   constant multiplicative factor, known  from the relation of Ising model and $Z_2$ percolation.

\section*{Appendix}
 In this Appendix, we examine the impact of particle density, $\rho $, on the phase separation transition. The RTP model under study involves three key parameters: $p $ and $\omega $, which govern the dynamics, and the conserved particle density $\rho $. In the phase-separated state, a region of high particle density ($\rho_> $) coexists with a region of low density ($\rho_< $). Therefore, the phase diagram in the $\rho$-$\omega $ plane for a fixed $p $ (or in the $\rho$-$p$ plane for a fixed $\omega $) is expected to display a first-order coexistence curve.
\begin{figure}[h]
\centering
\includegraphics[width=8cm,height=6 cm]{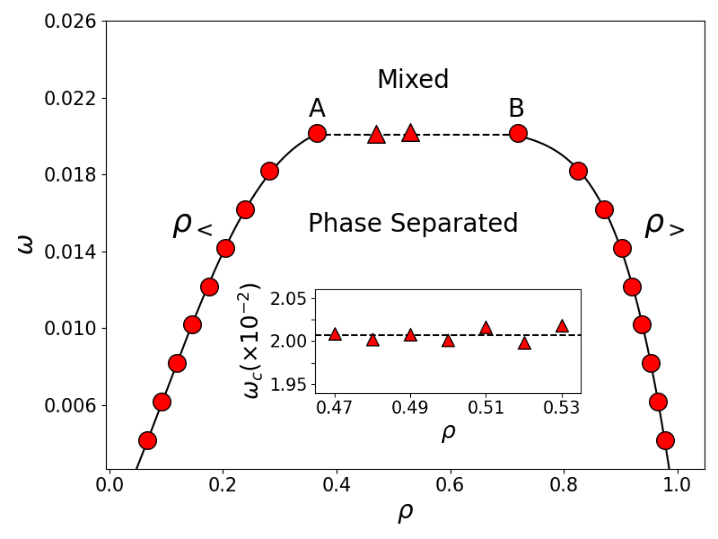}
\caption{ (Color online) Phase coexistence curve for  $p=0.235$ in a $6l\times 2l$ system.
The high-density $\rho_>$ is obtained  from  $l\times l$  subsystems
centered around the system’s center of mass, while the low-density 
$\rho_<$ is determined from subsystems shifted by $3l.$ A and B marks the maximum value of $\rho_<$ and the minimum value of $\rho_>$ that can be clearly resolved from the simulations. The inset  shows  $\omega_c$  for different $\rho,$ obtained from  the crossing point of  Binder cumulant $U_4$  in Eq. \eqref{eq:BC}. 
The values of $\omega_c$ remain nearly constant in the range $\rho\in(0.45,0.55)$ and align well with the expected coexistence line connecting A and B.
}
\label{fig:7}
\end{figure}

To determine the coexistence curve that separates the well-mixed phase from the phase-separated state, we follow the method proposed in Ref.~\cite{PhysRevE.98.030601} , where they studied phase transitions of active Brownian particles in two dimensions. We consider a rectangular lattice of size $6l \times 2l $ to encourage high-density clusters to align with the shorter axis. The particle density in two subsystem, each of size $l \times l $, centered around the system’s center of mass, provides estimates for $\rho_> $, while two subsystem shifted by 3$l $ provide estimates for $\rho_< $.

Figure~\ref{fig:7} illustrates the phase diagram for $p = 0.235 $ on a $192 \times 64 $ lattice ($l = 32 $), where $\rho_< $ and $\rho_> $ are indicated by filled circles. It is challenging to distinguish between the high- and low-density regimes when $\rho \approx \frac{1}{2} $.  In this regime ($0.45 < \rho < 0.55 $),  instead of calculating $\rho_<$ and $\rho_>$ we  try to  calculate the critical $\omega_c $  for different  $\rho$  from the crossing point of the Binder cumulant $U_4,$ defined in \eqref{eq:BC}.  
Resulting $\omega_c$  for  various $\rho $ are  presented  in the inset of Fig.~\ref{fig:7}. We observe that $\omega_c $ exhibits minimal variation within this density range, with differences on the order of $O(10^{-4}) $. Also, the values of $\omega_c$   compares  well  with the  line  that  joins  the points A and B  in coexistence curve. 

\begin{figure}
     \centering
     \includegraphics[width=4.25cm]{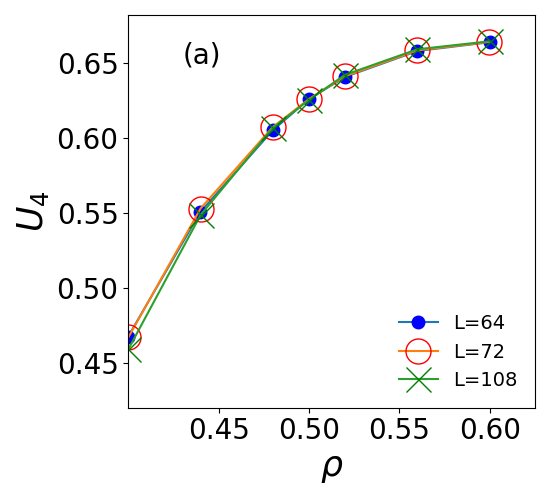}
     \includegraphics[width=4.25cm]{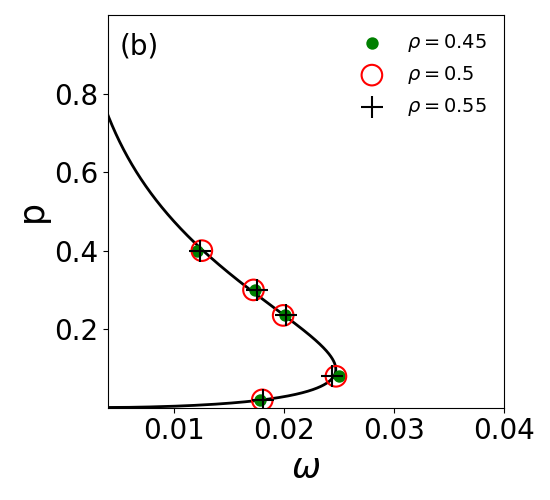} 
     \caption{ (Color online) (a)  $U_4$ vs $\rho$ for the critical point I ($p_c=0.235, \omega_c=0.02).$  Data for different $L$ merge with each other in a large  range  of  density, $\rho\in(0.45,0.55)$,  indicating that  the critical line is not affected   significantly  this density region. (b) The critical points  $(p_c,\omega_c)$ obtained  from the crossing points  of  $U_4$ for  $\rho=0.45$ (closed circle) and  $\rho=0.55$  ($+$ symbol) matches  closely  with  the  same obtained for $\rho=0.5$ (open circle and the guiding line). }
     \label{fig:bc_rho}
 \end{figure}

 Additionally, we attempt to estimate $\rho_c $  from  $U_4$  as  a  function of $\rho$  for different $L,$ keeping   $p,\omega$ fixed at  $p = 0.235 $ and $\omega = 0.02$,  which is  shown  in Fig.~\ref{fig:bc_rho}(a). The $U_4 $ values for three different system sizes $L $ coincide with each other over a wide density range and thus the crossing point could not be determined accurately; the critical density  $\rho_c$  lies  somewhere in  the range  $(0.45,0.55 $. This large uncertainty in  estimate of  $\rho_c$  is common  to find  in  phase separation transitions \cite{PhysRevLett.123.068002};  the flatness of the coexistence curve near its maximum  is responsible for this. 
In our study we  consider $\rho_c= \frac{1}{2}$ anticipating that  
the deviation of density  from its actual critical value  would  not affect the finite-size scaling or estimates of the critical exponents.
To validate this  further  we recalculated the phase diagram for $\rho = 0.45 $ and $\rho = 0.55 $, and compared it to the results for $\rho = 0.5 $ in Fig.~\ref{fig:bc_rho}(b); they match closely.

\section*{Acknowledgements}
The authors like to  acknowledge  helpful discussions with   Urna Basu and Indranil Mukherjee.

\bibliography{percoRTP_v2} 

\begin{thebibliography}{63}%
\makeatletter
\providecommand \@ifxundefined [1]{%
 \@ifx{#1\undefined}
}%
\providecommand \@ifnum [1]{%
 \ifnum #1\expandafter \@firstoftwo
 \else \expandafter \@secondoftwo
 \fi
}%
\providecommand \@ifx [1]{%
 \ifx #1\expandafter \@firstoftwo
 \else \expandafter \@secondoftwo
 \fi
}%
\providecommand \natexlab [1]{#1}%
\providecommand \enquote  [1]{``#1''}%
\providecommand \bibnamefont  [1]{#1}%
\providecommand \bibfnamefont [1]{#1}%
\providecommand \citenamefont [1]{#1}%
\providecommand \href@noop [0]{\@secondoftwo}%
\providecommand \href [0]{\begingroup \@sanitize@url \@href}%
\providecommand \@href[1]{\@@startlink{#1}\@@href}%
\providecommand \@@href[1]{\endgroup#1\@@endlink}%
\providecommand \@sanitize@url [0]{\catcode `\\12\catcode `\$12\catcode
  `\&12\catcode `\#12\catcode `\^12\catcode `\_12\catcode `\%12\relax}%
\providecommand \@@startlink[1]{}%
\providecommand \@@endlink[0]{}%
\providecommand \url  [0]{\begingroup\@sanitize@url \@url }%
\providecommand \@url [1]{\endgroup\@href {#1}{\urlprefix }}%
\providecommand \urlprefix  [0]{URL }%
\providecommand \Eprint [0]{\href }%
\providecommand \doibase [0]{http://dx.doi.org/}%
\providecommand \selectlanguage [0]{\@gobble}%
\providecommand \bibinfo  [0]{\@secondoftwo}%
\providecommand \bibfield  [0]{\@secondoftwo}%
\providecommand \translation [1]{[#1]}%
\providecommand \BibitemOpen [0]{}%
\providecommand \bibitemStop [0]{}%
\providecommand \bibitemNoStop [0]{.\EOS\space}%
\providecommand \EOS [0]{\spacefactor3000\relax}%
\providecommand \BibitemShut  [1]{\csname bibitem#1\endcsname}%
\let\auto@bib@innerbib\@empty
\bibitem [{\citenamefont
  {Ramaswamy}(2010)}]{annurev:/content/journals/10.1146/annurev-conmatphys-070909-104101}%
  \BibitemOpen
  \bibfield  {author} {\bibinfo {author} {\bibfnamefont {S.}~\bibnamefont
  {Ramaswamy}},\ }\href {\doibase
  https://doi.org/10.1146/annurev-conmatphys-070909-104101} {\bibfield
  {journal} {\bibinfo  {journal} {Annual Review of Condensed Matter Physics}\
  }\textbf {\bibinfo {volume} {1}},\ \bibinfo {pages} {323} (\bibinfo {year}
  {2010})}\BibitemShut {NoStop}%
\bibitem [{\citenamefont {Cates}(2012)}]{Cates_2012}%
  \BibitemOpen
  \bibfield  {author} {\bibinfo {author} {\bibfnamefont {M.~E.}\ \bibnamefont
  {Cates}},\ }\href {\doibase 10.1088/0034-4885/75/4/042601} {\bibfield
  {journal} {\bibinfo  {journal} {Reports on Progress in Physics}\ }\textbf
  {\bibinfo {volume} {75}},\ \bibinfo {pages} {042601} (\bibinfo {year}
  {2012})}\BibitemShut {NoStop}%
\bibitem [{\citenamefont {Marchetti}\ \emph {et~al.}(2013)\citenamefont
  {Marchetti}, \citenamefont {Joanny}, \citenamefont {Ramaswamy}, \citenamefont
  {Liverpool}, \citenamefont {Prost}, \citenamefont {Rao},\ and\ \citenamefont
  {Simha}}]{RevModPhys.85.1143}%
  \BibitemOpen
  \bibfield  {author} {\bibinfo {author} {\bibfnamefont {M.~C.}\ \bibnamefont
  {Marchetti}}, \bibinfo {author} {\bibfnamefont {J.~F.}\ \bibnamefont
  {Joanny}}, \bibinfo {author} {\bibfnamefont {S.}~\bibnamefont {Ramaswamy}},
  \bibinfo {author} {\bibfnamefont {T.~B.}\ \bibnamefont {Liverpool}}, \bibinfo
  {author} {\bibfnamefont {J.}~\bibnamefont {Prost}}, \bibinfo {author}
  {\bibfnamefont {M.}~\bibnamefont {Rao}}, \ and\ \bibinfo {author}
  {\bibfnamefont {R.~A.}\ \bibnamefont {Simha}},\ }\href {\doibase
  10.1103/RevModPhys.85.1143} {\bibfield  {journal} {\bibinfo  {journal} {Rev.
  Mod. Phys.}\ }\textbf {\bibinfo {volume} {85}},\ \bibinfo {pages} {1143}
  (\bibinfo {year} {2013})}\BibitemShut {NoStop}%
\bibitem [{\citenamefont {{De Magistris}}\ and\ \citenamefont
  {Marenduzzo}(2015)}]{DEMAGISTRIS201565}%
  \BibitemOpen
  \bibfield  {author} {\bibinfo {author} {\bibfnamefont {G.}~\bibnamefont {{De
  Magistris}}}\ and\ \bibinfo {author} {\bibfnamefont {D.}~\bibnamefont
  {Marenduzzo}},\ }\href {\doibase https://doi.org/10.1016/j.physa.2014.06.061}
  {\bibfield  {journal} {\bibinfo  {journal} {Physica A: Statistical Mechanics
  and its Applications}\ }\textbf {\bibinfo {volume} {418}},\ \bibinfo {pages}
  {65} (\bibinfo {year} {2015})},\ \bibinfo {note} {proceedings of the 13th
  International Summer School on Fundamental Problems in Statistical
  Physics}\BibitemShut {NoStop}%
\bibitem [{\citenamefont {Bechinger}\ \emph {et~al.}(2016)\citenamefont
  {Bechinger}, \citenamefont {Di~Leonardo}, \citenamefont {L\"owen},
  \citenamefont {Reichhardt}, \citenamefont {Volpe},\ and\ \citenamefont
  {Volpe}}]{RevModPhys.88.045006}%
  \BibitemOpen
  \bibfield  {author} {\bibinfo {author} {\bibfnamefont {C.}~\bibnamefont
  {Bechinger}}, \bibinfo {author} {\bibfnamefont {R.}~\bibnamefont
  {Di~Leonardo}}, \bibinfo {author} {\bibfnamefont {H.}~\bibnamefont
  {L\"owen}}, \bibinfo {author} {\bibfnamefont {C.}~\bibnamefont {Reichhardt}},
  \bibinfo {author} {\bibfnamefont {G.}~\bibnamefont {Volpe}}, \ and\ \bibinfo
  {author} {\bibfnamefont {G.}~\bibnamefont {Volpe}},\ }\href {\doibase
  10.1103/RevModPhys.88.045006} {\bibfield  {journal} {\bibinfo  {journal}
  {Rev. Mod. Phys.}\ }\textbf {\bibinfo {volume} {88}},\ \bibinfo {pages}
  {045006} (\bibinfo {year} {2016})}\BibitemShut {NoStop}%
\bibitem [{\citenamefont {Alert}\ and\ \citenamefont
  {Trepat}(2020)}]{annurev:/content/journals/10.1146/annurev-conmatphys-031218-013516}%
  \BibitemOpen
  \bibfield  {author} {\bibinfo {author} {\bibfnamefont {R.}~\bibnamefont
  {Alert}}\ and\ \bibinfo {author} {\bibfnamefont {X.}~\bibnamefont {Trepat}},\
  }\href {\doibase https://doi.org/10.1146/annurev-conmatphys-031218-013516}
  {\bibfield  {journal} {\bibinfo  {journal} {Annual Review of Condensed Matter
  Physics}\ }\textbf {\bibinfo {volume} {11}},\ \bibinfo {pages} {77} (\bibinfo
  {year} {2020})}\BibitemShut {NoStop}%
\bibitem [{\citenamefont {Ballerini}\ \emph {et~al.}(2008)\citenamefont
  {Ballerini}, \citenamefont {Cabibbo}, \citenamefont {Candelier},
  \citenamefont {Cavagna}, \citenamefont {Cisbani}, \citenamefont {Giardina},
  \citenamefont {Lecomte}, \citenamefont {Orlandi}, \citenamefont {Parisi},
  \citenamefont {Procaccini}, \citenamefont {Viale},\ and\ \citenamefont
  {Zdravkovic}}]{doi:10.1073/pnas.0711437105}%
  \BibitemOpen
  \bibfield  {author} {\bibinfo {author} {\bibfnamefont {M.}~\bibnamefont
  {Ballerini}}, \bibinfo {author} {\bibfnamefont {N.}~\bibnamefont {Cabibbo}},
  \bibinfo {author} {\bibfnamefont {R.}~\bibnamefont {Candelier}}, \bibinfo
  {author} {\bibfnamefont {A.}~\bibnamefont {Cavagna}}, \bibinfo {author}
  {\bibfnamefont {E.}~\bibnamefont {Cisbani}}, \bibinfo {author} {\bibfnamefont
  {I.}~\bibnamefont {Giardina}}, \bibinfo {author} {\bibfnamefont
  {V.}~\bibnamefont {Lecomte}}, \bibinfo {author} {\bibfnamefont
  {A.}~\bibnamefont {Orlandi}}, \bibinfo {author} {\bibfnamefont
  {G.}~\bibnamefont {Parisi}}, \bibinfo {author} {\bibfnamefont
  {A.}~\bibnamefont {Procaccini}}, \bibinfo {author} {\bibfnamefont
  {M.}~\bibnamefont {Viale}}, \ and\ \bibinfo {author} {\bibfnamefont
  {V.}~\bibnamefont {Zdravkovic}},\ }\href {\doibase 10.1073/pnas.0711437105}
  {\bibfield  {journal} {\bibinfo  {journal} {Proceedings of the National
  Academy of Sciences}\ }\textbf {\bibinfo {volume} {105}},\ \bibinfo {pages}
  {1232} (\bibinfo {year} {2008})},\ \Eprint
  {http://arxiv.org/abs/https://www.pnas.org/doi/pdf/10.1073/pnas.0711437105}
  {https://www.pnas.org/doi/pdf/10.1073/pnas.0711437105} \BibitemShut {NoStop}%
\bibitem [{\citenamefont {Ward}\ \emph {et~al.}(2008)\citenamefont {Ward},
  \citenamefont {Sumpter}, \citenamefont {Couzin}, \citenamefont {Hart},\ and\
  \citenamefont {Krause}}]{Ward2008-mw}%
  \BibitemOpen
  \bibfield  {author} {\bibinfo {author} {\bibfnamefont {A.~J.~W.}\
  \bibnamefont {Ward}}, \bibinfo {author} {\bibfnamefont {D.~J.~T.}\
  \bibnamefont {Sumpter}}, \bibinfo {author} {\bibfnamefont {I.~D.}\
  \bibnamefont {Couzin}}, \bibinfo {author} {\bibfnamefont {P.~J.~B.}\
  \bibnamefont {Hart}}, \ and\ \bibinfo {author} {\bibfnamefont
  {J.}~\bibnamefont {Krause}},\ }\href@noop {} {\bibfield  {journal} {\bibinfo
  {journal} {Proc. Natl. Acad. Sci. U. S. A.}\ }\textbf {\bibinfo {volume}
  {105}},\ \bibinfo {pages} {6948} (\bibinfo {year} {2008})}\BibitemShut
  {NoStop}%
\bibitem [{\citenamefont {Cavagna}\ \emph {et~al.}(2017)\citenamefont
  {Cavagna}, \citenamefont {Conti}, \citenamefont {Creato}, \citenamefont
  {Del~Castello}, \citenamefont {Giardina}, \citenamefont {Grigera},
  \citenamefont {Melillo}, \citenamefont {Parisi},\ and\ \citenamefont
  {Viale}}]{Cavagna2017}%
  \BibitemOpen
  \bibfield  {author} {\bibinfo {author} {\bibfnamefont {A.}~\bibnamefont
  {Cavagna}}, \bibinfo {author} {\bibfnamefont {D.}~\bibnamefont {Conti}},
  \bibinfo {author} {\bibfnamefont {C.}~\bibnamefont {Creato}}, \bibinfo
  {author} {\bibfnamefont {L.}~\bibnamefont {Del~Castello}}, \bibinfo {author}
  {\bibfnamefont {I.}~\bibnamefont {Giardina}}, \bibinfo {author}
  {\bibfnamefont {T.~S.}\ \bibnamefont {Grigera}}, \bibinfo {author}
  {\bibfnamefont {S.}~\bibnamefont {Melillo}}, \bibinfo {author} {\bibfnamefont
  {L.}~\bibnamefont {Parisi}}, \ and\ \bibinfo {author} {\bibfnamefont
  {M.}~\bibnamefont {Viale}},\ }\href {\doibase 10.1038/nphys4153} {\bibfield
  {journal} {\bibinfo  {journal} {Nature Physics}\ }\textbf {\bibinfo {volume}
  {13}},\ \bibinfo {pages} {914} (\bibinfo {year} {2017})}\BibitemShut
  {NoStop}%
\bibitem [{\citenamefont {Be'er}\ \emph {et~al.}(2020)\citenamefont {Be'er},
  \citenamefont {Ilkanaiv}, \citenamefont {Gross}, \citenamefont {Kearns},
  \citenamefont {Heidenreich}, \citenamefont {B{\"a}r},\ and\ \citenamefont
  {Ariel}}]{Beer2020}%
  \BibitemOpen
  \bibfield  {author} {\bibinfo {author} {\bibfnamefont {A.}~\bibnamefont
  {Be'er}}, \bibinfo {author} {\bibfnamefont {B.}~\bibnamefont {Ilkanaiv}},
  \bibinfo {author} {\bibfnamefont {R.}~\bibnamefont {Gross}}, \bibinfo
  {author} {\bibfnamefont {D.~B.}\ \bibnamefont {Kearns}}, \bibinfo {author}
  {\bibfnamefont {S.}~\bibnamefont {Heidenreich}}, \bibinfo {author}
  {\bibfnamefont {M.}~\bibnamefont {B{\"a}r}}, \ and\ \bibinfo {author}
  {\bibfnamefont {G.}~\bibnamefont {Ariel}},\ }\href {\doibase
  10.1038/s42005-020-0327-1} {\bibfield  {journal} {\bibinfo  {journal}
  {Communications Physics}\ }\textbf {\bibinfo {volume} {3}},\ \bibinfo {pages}
  {66} (\bibinfo {year} {2020})}\BibitemShut {NoStop}%
\bibitem [{\citenamefont {Peruani}\ \emph {et~al.}(2012)\citenamefont
  {Peruani}, \citenamefont {Starru\ss{}}, \citenamefont {Jakovljevic},
  \citenamefont {S\o{}gaard-Andersen}, \citenamefont {Deutsch},\ and\
  \citenamefont {B\"ar}}]{PhysRevLett.108.098102}%
  \BibitemOpen
  \bibfield  {author} {\bibinfo {author} {\bibfnamefont {F.}~\bibnamefont
  {Peruani}}, \bibinfo {author} {\bibfnamefont {J.}~\bibnamefont
  {Starru\ss{}}}, \bibinfo {author} {\bibfnamefont {V.}~\bibnamefont
  {Jakovljevic}}, \bibinfo {author} {\bibfnamefont {L.}~\bibnamefont
  {S\o{}gaard-Andersen}}, \bibinfo {author} {\bibfnamefont {A.}~\bibnamefont
  {Deutsch}}, \ and\ \bibinfo {author} {\bibfnamefont {M.}~\bibnamefont
  {B\"ar}},\ }\href {\doibase 10.1103/PhysRevLett.108.098102} {\bibfield
  {journal} {\bibinfo  {journal} {Phys. Rev. Lett.}\ }\textbf {\bibinfo
  {volume} {108}},\ \bibinfo {pages} {098102} (\bibinfo {year}
  {2012})}\BibitemShut {NoStop}%
\bibitem [{\citenamefont {Berg}(2004)}]{berg2003ecoli}%
  \BibitemOpen
  \bibinfo {editor} {\bibfnamefont {H.~C.}\ \bibnamefont {Berg}},\ ed.,\ \href
  {\doibase 10.1007/b97370} {\emph {\bibinfo {title} {E. coli in Motion}}},\
  \bibinfo {edition} {1st}\ ed.,\ Biological and Medical Physics, Biomedical
  Engineering\ (\bibinfo  {publisher} {Springer New York, NY},\ \bibinfo {year}
  {2004})\ pp.\ \bibinfo {pages} {XII, 134},\ \bibinfo {note} {published: 01
  October 2003 (Hardcover), 12 December 2011 (Softcover), 11 January 2008
  (eBook)}\BibitemShut {NoStop}%
\bibitem [{\citenamefont {Polin}\ \emph {et~al.}(2009)\citenamefont {Polin},
  \citenamefont {Tuval}, \citenamefont {Drescher}, \citenamefont {Gollub},\
  and\ \citenamefont {Goldstein}}]{doi:10.1126/science.1172667}%
  \BibitemOpen
  \bibfield  {author} {\bibinfo {author} {\bibfnamefont {M.}~\bibnamefont
  {Polin}}, \bibinfo {author} {\bibfnamefont {I.}~\bibnamefont {Tuval}},
  \bibinfo {author} {\bibfnamefont {K.}~\bibnamefont {Drescher}}, \bibinfo
  {author} {\bibfnamefont {J.~P.}\ \bibnamefont {Gollub}}, \ and\ \bibinfo
  {author} {\bibfnamefont {R.~E.}\ \bibnamefont {Goldstein}},\ }\href {\doibase
  10.1126/science.1172667} {\bibfield  {journal} {\bibinfo  {journal}
  {Science}\ }\textbf {\bibinfo {volume} {325}},\ \bibinfo {pages} {487}
  (\bibinfo {year} {2009})},\ \Eprint
  {http://arxiv.org/abs/https://www.science.org/doi/pdf/10.1126/science.1172667}
  {https://www.science.org/doi/pdf/10.1126/science.1172667} \BibitemShut
  {NoStop}%
\bibitem [{\citenamefont {Tailleur}\ and\ \citenamefont
  {Cates}(2008)}]{PhysRevLett.100.218103}%
  \BibitemOpen
  \bibfield  {author} {\bibinfo {author} {\bibfnamefont {J.}~\bibnamefont
  {Tailleur}}\ and\ \bibinfo {author} {\bibfnamefont {M.~E.}\ \bibnamefont
  {Cates}},\ }\href {\doibase 10.1103/PhysRevLett.100.218103} {\bibfield
  {journal} {\bibinfo  {journal} {Phys. Rev. Lett.}\ }\textbf {\bibinfo
  {volume} {100}},\ \bibinfo {pages} {218103} (\bibinfo {year}
  {2008})}\BibitemShut {NoStop}%
\bibitem [{\citenamefont {Cates}\ and\ \citenamefont
  {Tailleur}(2013)}]{Cates_2013}%
  \BibitemOpen
  \bibfield  {author} {\bibinfo {author} {\bibfnamefont {M.~E.}\ \bibnamefont
  {Cates}}\ and\ \bibinfo {author} {\bibfnamefont {J.}~\bibnamefont
  {Tailleur}},\ }\href {\doibase 10.1209/0295-5075/101/20010} {\bibfield
  {journal} {\bibinfo  {journal} {Europhysics Letters}\ }\textbf {\bibinfo
  {volume} {101}},\ \bibinfo {pages} {20010} (\bibinfo {year}
  {2013})}\BibitemShut {NoStop}%
\bibitem [{\citenamefont {Cates}\ and\ \citenamefont
  {Tailleur}(2015)}]{annurev:/content/journals/10.1146/annurev-conmatphys-031214-014710}%
  \BibitemOpen
  \bibfield  {author} {\bibinfo {author} {\bibfnamefont {M.~E.}\ \bibnamefont
  {Cates}}\ and\ \bibinfo {author} {\bibfnamefont {J.}~\bibnamefont
  {Tailleur}},\ }\href {\doibase
  https://doi.org/10.1146/annurev-conmatphys-031214-014710} {\bibfield
  {journal} {\bibinfo  {journal} {Annual Review of Condensed Matter Physics}\
  }\textbf {\bibinfo {volume} {6}},\ \bibinfo {pages} {219} (\bibinfo {year}
  {2015})}\BibitemShut {NoStop}%
\bibitem [{\citenamefont {Fily}\ and\ \citenamefont
  {Marchetti}(2012)}]{PhysRevLett.108.235702}%
  \BibitemOpen
  \bibfield  {author} {\bibinfo {author} {\bibfnamefont {Y.}~\bibnamefont
  {Fily}}\ and\ \bibinfo {author} {\bibfnamefont {M.~C.}\ \bibnamefont
  {Marchetti}},\ }\href {\doibase 10.1103/PhysRevLett.108.235702} {\bibfield
  {journal} {\bibinfo  {journal} {Phys. Rev. Lett.}\ }\textbf {\bibinfo
  {volume} {108}},\ \bibinfo {pages} {235702} (\bibinfo {year}
  {2012})}\BibitemShut {NoStop}%
\bibitem [{\citenamefont {Fily}\ \emph {et~al.}(2014)\citenamefont {Fily},
  \citenamefont {Henkes},\ and\ \citenamefont {Marchetti}}]{C3SM52469H}%
  \BibitemOpen
  \bibfield  {author} {\bibinfo {author} {\bibfnamefont {Y.}~\bibnamefont
  {Fily}}, \bibinfo {author} {\bibfnamefont {S.}~\bibnamefont {Henkes}}, \ and\
  \bibinfo {author} {\bibfnamefont {M.~C.}\ \bibnamefont {Marchetti}},\ }\href
  {\doibase 10.1039/C3SM52469H} {\bibfield  {journal} {\bibinfo  {journal}
  {Soft Matter}\ }\textbf {\bibinfo {volume} {10}},\ \bibinfo {pages} {2132}
  (\bibinfo {year} {2014})}\BibitemShut {NoStop}%
\bibitem [{\citenamefont {Redner}\ \emph {et~al.}(2013)\citenamefont {Redner},
  \citenamefont {Hagan},\ and\ \citenamefont
  {Baskaran}}]{PhysRevLett.110.055701}%
  \BibitemOpen
  \bibfield  {author} {\bibinfo {author} {\bibfnamefont {G.~S.}\ \bibnamefont
  {Redner}}, \bibinfo {author} {\bibfnamefont {M.~F.}\ \bibnamefont {Hagan}}, \
  and\ \bibinfo {author} {\bibfnamefont {A.}~\bibnamefont {Baskaran}},\ }\href
  {\doibase 10.1103/PhysRevLett.110.055701} {\bibfield  {journal} {\bibinfo
  {journal} {Phys. Rev. Lett.}\ }\textbf {\bibinfo {volume} {110}},\ \bibinfo
  {pages} {055701} (\bibinfo {year} {2013})}\BibitemShut {NoStop}%
\bibitem [{\citenamefont {Stenhammar}\ \emph {et~al.}(2013)\citenamefont
  {Stenhammar}, \citenamefont {Tiribocchi}, \citenamefont {Allen},
  \citenamefont {Marenduzzo},\ and\ \citenamefont
  {Cates}}]{PhysRevLett.111.145702}%
  \BibitemOpen
  \bibfield  {author} {\bibinfo {author} {\bibfnamefont {J.}~\bibnamefont
  {Stenhammar}}, \bibinfo {author} {\bibfnamefont {A.}~\bibnamefont
  {Tiribocchi}}, \bibinfo {author} {\bibfnamefont {R.~J.}\ \bibnamefont
  {Allen}}, \bibinfo {author} {\bibfnamefont {D.}~\bibnamefont {Marenduzzo}}, \
  and\ \bibinfo {author} {\bibfnamefont {M.~E.}\ \bibnamefont {Cates}},\ }\href
  {\doibase 10.1103/PhysRevLett.111.145702} {\bibfield  {journal} {\bibinfo
  {journal} {Phys. Rev. Lett.}\ }\textbf {\bibinfo {volume} {111}},\ \bibinfo
  {pages} {145702} (\bibinfo {year} {2013})}\BibitemShut {NoStop}%
\bibitem [{\citenamefont {Gonnella}\ \emph {et~al.}(2014)\citenamefont
  {Gonnella}, \citenamefont {Lamura},\ and\ \citenamefont
  {Suma}}]{doi:10.1142/S0129183114410046}%
  \BibitemOpen
  \bibfield  {author} {\bibinfo {author} {\bibfnamefont {G.}~\bibnamefont
  {Gonnella}}, \bibinfo {author} {\bibfnamefont {A.}~\bibnamefont {Lamura}}, \
  and\ \bibinfo {author} {\bibfnamefont {A.}~\bibnamefont {Suma}},\ }\href
  {\doibase 10.1142/S0129183114410046} {\bibfield  {journal} {\bibinfo
  {journal} {International Journal of Modern Physics C}\ }\textbf {\bibinfo
  {volume} {25}},\ \bibinfo {pages} {1441004} (\bibinfo {year} {2014})},\
  \Eprint {http://arxiv.org/abs/https://doi.org/10.1142/S0129183114410046}
  {https://doi.org/10.1142/S0129183114410046} \BibitemShut {NoStop}%
\bibitem [{\citenamefont {Suma}\ \emph {et~al.}(2014)\citenamefont {Suma},
  \citenamefont {Gonnella}, \citenamefont {Marenduzzo},\ and\ \citenamefont
  {Orlandini}}]{Suma_2014}%
  \BibitemOpen
  \bibfield  {author} {\bibinfo {author} {\bibfnamefont {A.}~\bibnamefont
  {Suma}}, \bibinfo {author} {\bibfnamefont {G.}~\bibnamefont {Gonnella}},
  \bibinfo {author} {\bibfnamefont {D.}~\bibnamefont {Marenduzzo}}, \ and\
  \bibinfo {author} {\bibfnamefont {E.}~\bibnamefont {Orlandini}},\ }\href
  {\doibase 10.1209/0295-5075/108/56004} {\bibfield  {journal} {\bibinfo
  {journal} {Europhysics Letters}\ }\textbf {\bibinfo {volume} {108}},\
  \bibinfo {pages} {56004} (\bibinfo {year} {2014})}\BibitemShut {NoStop}%
\bibitem [{\citenamefont {Levis}\ and\ \citenamefont
  {Berthier}(2014)}]{PhysRevE.89.062301}%
  \BibitemOpen
  \bibfield  {author} {\bibinfo {author} {\bibfnamefont {D.}~\bibnamefont
  {Levis}}\ and\ \bibinfo {author} {\bibfnamefont {L.}~\bibnamefont
  {Berthier}},\ }\href {\doibase 10.1103/PhysRevE.89.062301} {\bibfield
  {journal} {\bibinfo  {journal} {Phys. Rev. E}\ }\textbf {\bibinfo {volume}
  {89}},\ \bibinfo {pages} {062301} (\bibinfo {year} {2014})}\BibitemShut
  {NoStop}%
\bibitem [{\citenamefont {Wittkowski}\ \emph {et~al.}(2014)\citenamefont
  {Wittkowski}, \citenamefont {Tiribocchi}, \citenamefont {Stenhammar},
  \citenamefont {Allen}, \citenamefont {Marenduzzo},\ and\ \citenamefont
  {Cates}}]{Wittkowski2014}%
  \BibitemOpen
  \bibfield  {author} {\bibinfo {author} {\bibfnamefont {R.}~\bibnamefont
  {Wittkowski}}, \bibinfo {author} {\bibfnamefont {A.}~\bibnamefont
  {Tiribocchi}}, \bibinfo {author} {\bibfnamefont {J.}~\bibnamefont
  {Stenhammar}}, \bibinfo {author} {\bibfnamefont {R.~J.}\ \bibnamefont
  {Allen}}, \bibinfo {author} {\bibfnamefont {D.}~\bibnamefont {Marenduzzo}}, \
  and\ \bibinfo {author} {\bibfnamefont {M.~E.}\ \bibnamefont {Cates}},\ }\href
  {\doibase 10.1038/ncomms5351} {\bibfield  {journal} {\bibinfo  {journal}
  {Nature Communications}\ }\textbf {\bibinfo {volume} {5}},\ \bibinfo {pages}
  {4351} (\bibinfo {year} {2014})}\BibitemShut {NoStop}%
\bibitem [{\citenamefont {Kourbane-Houssene}\ \emph {et~al.}(2018)\citenamefont
  {Kourbane-Houssene}, \citenamefont {Erignoux}, \citenamefont {Bodineau},\
  and\ \citenamefont {Tailleur}}]{PhysRevLett.120.268003}%
  \BibitemOpen
  \bibfield  {author} {\bibinfo {author} {\bibfnamefont {M.}~\bibnamefont
  {Kourbane-Houssene}}, \bibinfo {author} {\bibfnamefont {C.}~\bibnamefont
  {Erignoux}}, \bibinfo {author} {\bibfnamefont {T.}~\bibnamefont {Bodineau}},
  \ and\ \bibinfo {author} {\bibfnamefont {J.}~\bibnamefont {Tailleur}},\
  }\href {\doibase 10.1103/PhysRevLett.120.268003} {\bibfield  {journal}
  {\bibinfo  {journal} {Phys. Rev. Lett.}\ }\textbf {\bibinfo {volume} {120}},\
  \bibinfo {pages} {268003} (\bibinfo {year} {2018})}\BibitemShut {NoStop}%
\bibitem [{\citenamefont {Bialké}\ \emph {et~al.}(2013)\citenamefont
  {Bialké}, \citenamefont {Löwen},\ and\ \citenamefont
  {Speck}}]{Bialke_2013}%
  \BibitemOpen
  \bibfield  {author} {\bibinfo {author} {\bibfnamefont {J.}~\bibnamefont
  {Bialké}}, \bibinfo {author} {\bibfnamefont {H.}~\bibnamefont {Löwen}}, \
  and\ \bibinfo {author} {\bibfnamefont {T.}~\bibnamefont {Speck}},\ }\href
  {\doibase 10.1209/0295-5075/103/30008} {\bibfield  {journal} {\bibinfo
  {journal} {Europhysics Letters}\ }\textbf {\bibinfo {volume} {103}},\
  \bibinfo {pages} {30008} (\bibinfo {year} {2013})}\BibitemShut {NoStop}%
\bibitem [{\citenamefont {Schweitzer}(2018)}]{Schweitzer_2019}%
  \BibitemOpen
  \bibfield  {author} {\bibinfo {author} {\bibfnamefont {F.}~\bibnamefont
  {Schweitzer}},\ }\href {\doibase 10.1088/1361-6404/aaeb63} {\bibfield
  {journal} {\bibinfo  {journal} {European Journal of Physics}\ }\textbf
  {\bibinfo {volume} {40}},\ \bibinfo {pages} {014003} (\bibinfo {year}
  {2018})}\BibitemShut {NoStop}%
\bibitem [{\citenamefont {Ziepke}\ \emph {et~al.}(2022)\citenamefont {Ziepke},
  \citenamefont {Maryshev}, \citenamefont {Aranson},\ and\ \citenamefont
  {Frey}}]{Ziepke2022}%
  \BibitemOpen
  \bibfield  {author} {\bibinfo {author} {\bibfnamefont {A.}~\bibnamefont
  {Ziepke}}, \bibinfo {author} {\bibfnamefont {I.}~\bibnamefont {Maryshev}},
  \bibinfo {author} {\bibfnamefont {I.~S.}\ \bibnamefont {Aranson}}, \ and\
  \bibinfo {author} {\bibfnamefont {E.}~\bibnamefont {Frey}},\ }\href {\doibase
  10.1038/s41467-022-34484-2} {\bibfield  {journal} {\bibinfo  {journal}
  {Nature Communications}\ }\textbf {\bibinfo {volume} {13}},\ \bibinfo {pages}
  {6727} (\bibinfo {year} {2022})}\BibitemShut {NoStop}%
\bibitem [{\citenamefont {Thompson}\ \emph {et~al.}(2011)\citenamefont
  {Thompson}, \citenamefont {Tailleur}, \citenamefont {Cates},\ and\
  \citenamefont {Blythe}}]{Thompson_2011}%
  \BibitemOpen
  \bibfield  {author} {\bibinfo {author} {\bibfnamefont {A.~G.}\ \bibnamefont
  {Thompson}}, \bibinfo {author} {\bibfnamefont {J.}~\bibnamefont {Tailleur}},
  \bibinfo {author} {\bibfnamefont {M.~E.}\ \bibnamefont {Cates}}, \ and\
  \bibinfo {author} {\bibfnamefont {R.~A.}\ \bibnamefont {Blythe}},\ }\href
  {\doibase 10.1088/1742-5468/2011/02/P02029} {\bibfield  {journal} {\bibinfo
  {journal} {Journal of Statistical Mechanics: Theory and Experiment}\ }\textbf
  {\bibinfo {volume} {2011}},\ \bibinfo {pages} {P02029} (\bibinfo {year}
  {2011})}\BibitemShut {NoStop}%
\bibitem [{\citenamefont {Slowman}\ \emph {et~al.}(2016)\citenamefont
  {Slowman}, \citenamefont {Evans},\ and\ \citenamefont
  {Blythe}}]{PhysRevLett.116.218101}%
  \BibitemOpen
  \bibfield  {author} {\bibinfo {author} {\bibfnamefont {A.~B.}\ \bibnamefont
  {Slowman}}, \bibinfo {author} {\bibfnamefont {M.~R.}\ \bibnamefont {Evans}},
  \ and\ \bibinfo {author} {\bibfnamefont {R.~A.}\ \bibnamefont {Blythe}},\
  }\href {\doibase 10.1103/PhysRevLett.116.218101} {\bibfield  {journal}
  {\bibinfo  {journal} {Phys. Rev. Lett.}\ }\textbf {\bibinfo {volume} {116}},\
  \bibinfo {pages} {218101} (\bibinfo {year} {2016})}\BibitemShut {NoStop}%
\bibitem [{\citenamefont {Mallmin}\ \emph {et~al.}(2019)\citenamefont
  {Mallmin}, \citenamefont {Blythe},\ and\ \citenamefont
  {Evans}}]{Mallmin_2019}%
  \BibitemOpen
  \bibfield  {author} {\bibinfo {author} {\bibfnamefont {E.}~\bibnamefont
  {Mallmin}}, \bibinfo {author} {\bibfnamefont {R.~A.}\ \bibnamefont {Blythe}},
  \ and\ \bibinfo {author} {\bibfnamefont {M.~R.}\ \bibnamefont {Evans}},\
  }\href {\doibase 10.1088/1742-5468/aaf631} {\bibfield  {journal} {\bibinfo
  {journal} {Journal of Statistical Mechanics: Theory and Experiment}\ }\textbf
  {\bibinfo {volume} {2019}},\ \bibinfo {pages} {013204} (\bibinfo {year}
  {2019})}\BibitemShut {NoStop}%
\bibitem [{\citenamefont {Dandekar}\ \emph {et~al.}(2020)\citenamefont
  {Dandekar}, \citenamefont {Chakraborti},\ and\ \citenamefont
  {Rajesh}}]{PhysRevE.102.062111}%
  \BibitemOpen
  \bibfield  {author} {\bibinfo {author} {\bibfnamefont {R.}~\bibnamefont
  {Dandekar}}, \bibinfo {author} {\bibfnamefont {S.}~\bibnamefont
  {Chakraborti}}, \ and\ \bibinfo {author} {\bibfnamefont {R.}~\bibnamefont
  {Rajesh}},\ }\href {\doibase 10.1103/PhysRevE.102.062111} {\bibfield
  {journal} {\bibinfo  {journal} {Phys. Rev. E}\ }\textbf {\bibinfo {volume}
  {102}},\ \bibinfo {pages} {062111} (\bibinfo {year} {2020})}\BibitemShut
  {NoStop}%
\bibitem [{\citenamefont {Mukherjee}\ \emph {et~al.}(2023)\citenamefont
  {Mukherjee}, \citenamefont {Raghu},\ and\ \citenamefont
  {Mohanty}}]{10.21468/SciPostPhys.14.6.165}%
  \BibitemOpen
  \bibfield  {author} {\bibinfo {author} {\bibfnamefont {I.}~\bibnamefont
  {Mukherjee}}, \bibinfo {author} {\bibfnamefont {A.}~\bibnamefont {Raghu}}, \
  and\ \bibinfo {author} {\bibfnamefont {P.~K.}\ \bibnamefont {Mohanty}},\
  }\href {\doibase 10.21468/SciPostPhys.14.6.165} {\bibfield  {journal}
  {\bibinfo  {journal} {SciPost Phys.}\ }\textbf {\bibinfo {volume} {14}},\
  \bibinfo {pages} {165} (\bibinfo {year} {2023})}\BibitemShut {NoStop}%
\bibitem [{\citenamefont {Palacci}\ \emph {et~al.}(2010)\citenamefont
  {Palacci}, \citenamefont {Ab\'ecassis}, \citenamefont {Cottin-Bizonne},
  \citenamefont {Ybert},\ and\ \citenamefont
  {Bocquet}}]{PhysRevLett.104.138302}%
  \BibitemOpen
  \bibfield  {author} {\bibinfo {author} {\bibfnamefont {J.}~\bibnamefont
  {Palacci}}, \bibinfo {author} {\bibfnamefont {B.}~\bibnamefont
  {Ab\'ecassis}}, \bibinfo {author} {\bibfnamefont {C.}~\bibnamefont
  {Cottin-Bizonne}}, \bibinfo {author} {\bibfnamefont {C.}~\bibnamefont
  {Ybert}}, \ and\ \bibinfo {author} {\bibfnamefont {L.}~\bibnamefont
  {Bocquet}},\ }\href {\doibase 10.1103/PhysRevLett.104.138302} {\bibfield
  {journal} {\bibinfo  {journal} {Phys. Rev. Lett.}\ }\textbf {\bibinfo
  {volume} {104}},\ \bibinfo {pages} {138302} (\bibinfo {year}
  {2010})}\BibitemShut {NoStop}%
\bibitem [{\citenamefont {Palacci}\ \emph {et~al.}(2014)\citenamefont
  {Palacci}, \citenamefont {Sacanna}, \citenamefont {Kim}, \citenamefont {Yi},
  \citenamefont {Pine},\ and\ \citenamefont
  {Chaikin}}]{doi:10.1098/rsta.2013.0372}%
  \BibitemOpen
  \bibfield  {author} {\bibinfo {author} {\bibfnamefont {J.}~\bibnamefont
  {Palacci}}, \bibinfo {author} {\bibfnamefont {S.}~\bibnamefont {Sacanna}},
  \bibinfo {author} {\bibfnamefont {S.-H.}\ \bibnamefont {Kim}}, \bibinfo
  {author} {\bibfnamefont {G.-R.}\ \bibnamefont {Yi}}, \bibinfo {author}
  {\bibfnamefont {D.~J.}\ \bibnamefont {Pine}}, \ and\ \bibinfo {author}
  {\bibfnamefont {P.~M.}\ \bibnamefont {Chaikin}},\ }\href {\doibase
  10.1098/rsta.2013.0372} {\bibfield  {journal} {\bibinfo  {journal}
  {Philosophical Transactions of the Royal Society A: Mathematical, Physical
  and Engineering Sciences}\ }\textbf {\bibinfo {volume} {372}},\ \bibinfo
  {pages} {20130372} (\bibinfo {year} {2014})}\BibitemShut {NoStop}%
\bibitem [{\citenamefont {Kushwaha}\ \emph {et~al.}(2023)\citenamefont
  {Kushwaha}, \citenamefont {Semwal}, \citenamefont {Maity}, \citenamefont
  {Mishra},\ and\ \citenamefont {Chikkadi}}]{PhysRevE.108.034603}%
  \BibitemOpen
  \bibfield  {author} {\bibinfo {author} {\bibfnamefont {P.}~\bibnamefont
  {Kushwaha}}, \bibinfo {author} {\bibfnamefont {V.}~\bibnamefont {Semwal}},
  \bibinfo {author} {\bibfnamefont {S.}~\bibnamefont {Maity}}, \bibinfo
  {author} {\bibfnamefont {S.}~\bibnamefont {Mishra}}, \ and\ \bibinfo {author}
  {\bibfnamefont {V.}~\bibnamefont {Chikkadi}},\ }\href {\doibase
  10.1103/PhysRevE.108.034603} {\bibfield  {journal} {\bibinfo  {journal}
  {Phys. Rev. E}\ }\textbf {\bibinfo {volume} {108}},\ \bibinfo {pages}
  {034603} (\bibinfo {year} {2023})}\BibitemShut {NoStop}%
\bibitem [{\citenamefont {Kushwaha}\ \emph {et~al.}(2024)\citenamefont
  {Kushwaha}, \citenamefont {Maity}, \citenamefont {Menon}, \citenamefont
  {Chelakkot},\ and\ \citenamefont {Chikkadi}}]{D4SM00305E}%
  \BibitemOpen
  \bibfield  {author} {\bibinfo {author} {\bibfnamefont {P.}~\bibnamefont
  {Kushwaha}}, \bibinfo {author} {\bibfnamefont {S.}~\bibnamefont {Maity}},
  \bibinfo {author} {\bibfnamefont {A.}~\bibnamefont {Menon}}, \bibinfo
  {author} {\bibfnamefont {R.}~\bibnamefont {Chelakkot}}, \ and\ \bibinfo
  {author} {\bibfnamefont {V.}~\bibnamefont {Chikkadi}},\ }\href {\doibase
  10.1039/D4SM00305E} {\bibfield  {journal} {\bibinfo  {journal} {Soft Matter}\
  }\textbf {\bibinfo {volume} {20}},\ \bibinfo {pages} {4699} (\bibinfo {year}
  {2024})}\BibitemShut {NoStop}%
\bibitem [{\citenamefont {Soto}\ and\ \citenamefont
  {Golestanian}(2014)}]{PhysRevE.89.012706}%
  \BibitemOpen
  \bibfield  {author} {\bibinfo {author} {\bibfnamefont {R.}~\bibnamefont
  {Soto}}\ and\ \bibinfo {author} {\bibfnamefont {R.}~\bibnamefont
  {Golestanian}},\ }\href {\doibase 10.1103/PhysRevE.89.012706} {\bibfield
  {journal} {\bibinfo  {journal} {Phys. Rev. E}\ }\textbf {\bibinfo {volume}
  {89}},\ \bibinfo {pages} {012706} (\bibinfo {year} {2014})}\BibitemShut
  {NoStop}%
\bibitem [{\citenamefont {Whitelam}\ \emph {et~al.}(2018)\citenamefont
  {Whitelam}, \citenamefont {Klymko},\ and\ \citenamefont
  {Mandal}}]{10.1063/1.5023403}%
  \BibitemOpen
  \bibfield  {author} {\bibinfo {author} {\bibfnamefont {S.}~\bibnamefont
  {Whitelam}}, \bibinfo {author} {\bibfnamefont {K.}~\bibnamefont {Klymko}}, \
  and\ \bibinfo {author} {\bibfnamefont {D.}~\bibnamefont {Mandal}},\ }\href
  {\doibase 10.1063/1.5023403} {\bibfield  {journal} {\bibinfo  {journal} {The
  Journal of Chemical Physics}\ }\textbf {\bibinfo {volume} {148}},\ \bibinfo
  {pages} {154902} (\bibinfo {year} {2018})},\ \Eprint
  {http://arxiv.org/abs/https://pubs.aip.org/aip/jcp/article-pdf/doi/10.1063/1.5023403/13854802/154902\_1\_online.pdf}
  {https://pubs.aip.org/aip/jcp/article-pdf/doi/10.1063/1.5023403/13854802/154902\_1\_online.pdf}
  \BibitemShut {NoStop}%
\bibitem [{\citenamefont {Solon}\ and\ \citenamefont
  {Tailleur}(2015)}]{PhysRevE.92.042119}%
  \BibitemOpen
  \bibfield  {author} {\bibinfo {author} {\bibfnamefont {A.~P.}\ \bibnamefont
  {Solon}}\ and\ \bibinfo {author} {\bibfnamefont {J.}~\bibnamefont
  {Tailleur}},\ }\href {\doibase 10.1103/PhysRevE.92.042119} {\bibfield
  {journal} {\bibinfo  {journal} {Phys. Rev. E}\ }\textbf {\bibinfo {volume}
  {92}},\ \bibinfo {pages} {042119} (\bibinfo {year} {2015})}\BibitemShut
  {NoStop}%
\bibitem [{\citenamefont {Sep\'ulveda}\ and\ \citenamefont
  {Soto}(2016)}]{PhysRevE.94.022603}%
  \BibitemOpen
  \bibfield  {author} {\bibinfo {author} {\bibfnamefont {N.}~\bibnamefont
  {Sep\'ulveda}}\ and\ \bibinfo {author} {\bibfnamefont {R.}~\bibnamefont
  {Soto}},\ }\href {\doibase 10.1103/PhysRevE.94.022603} {\bibfield  {journal}
  {\bibinfo  {journal} {Phys. Rev. E}\ }\textbf {\bibinfo {volume} {94}},\
  \bibinfo {pages} {022603} (\bibinfo {year} {2016})}\BibitemShut {NoStop}%
\bibitem [{\citenamefont {Siebert}\ \emph {et~al.}(2018)\citenamefont
  {Siebert}, \citenamefont {Dittrich}, \citenamefont {Schmid}, \citenamefont
  {Binder}, \citenamefont {Speck},\ and\ \citenamefont
  {Virnau}}]{PhysRevE.98.030601}%
  \BibitemOpen
  \bibfield  {author} {\bibinfo {author} {\bibfnamefont {J.~T.}\ \bibnamefont
  {Siebert}}, \bibinfo {author} {\bibfnamefont {F.}~\bibnamefont {Dittrich}},
  \bibinfo {author} {\bibfnamefont {F.}~\bibnamefont {Schmid}}, \bibinfo
  {author} {\bibfnamefont {K.}~\bibnamefont {Binder}}, \bibinfo {author}
  {\bibfnamefont {T.}~\bibnamefont {Speck}}, \ and\ \bibinfo {author}
  {\bibfnamefont {P.}~\bibnamefont {Virnau}},\ }\href {\doibase
  10.1103/PhysRevE.98.030601} {\bibfield  {journal} {\bibinfo  {journal} {Phys.
  Rev. E}\ }\textbf {\bibinfo {volume} {98}},\ \bibinfo {pages} {030601}
  (\bibinfo {year} {2018})}\BibitemShut {NoStop}%
\bibitem [{\citenamefont {Partridge}\ and\ \citenamefont
  {Lee}(2019)}]{PhysRevLett.123.068002}%
  \BibitemOpen
  \bibfield  {author} {\bibinfo {author} {\bibfnamefont {B.}~\bibnamefont
  {Partridge}}\ and\ \bibinfo {author} {\bibfnamefont {C.~F.}\ \bibnamefont
  {Lee}},\ }\href {\doibase 10.1103/PhysRevLett.123.068002} {\bibfield
  {journal} {\bibinfo  {journal} {Phys. Rev. Lett.}\ }\textbf {\bibinfo
  {volume} {123}},\ \bibinfo {pages} {068002} (\bibinfo {year}
  {2019})}\BibitemShut {NoStop}%
\bibitem [{\citenamefont {Maggi}\ \emph {et~al.}(2021)\citenamefont {Maggi},
  \citenamefont {Paoluzzi}, \citenamefont {Crisanti}, \citenamefont
  {Zaccarelli},\ and\ \citenamefont {Gnan}}]{D0SM02162H}%
  \BibitemOpen
  \bibfield  {author} {\bibinfo {author} {\bibfnamefont {C.}~\bibnamefont
  {Maggi}}, \bibinfo {author} {\bibfnamefont {M.}~\bibnamefont {Paoluzzi}},
  \bibinfo {author} {\bibfnamefont {A.}~\bibnamefont {Crisanti}}, \bibinfo
  {author} {\bibfnamefont {E.}~\bibnamefont {Zaccarelli}}, \ and\ \bibinfo
  {author} {\bibfnamefont {N.}~\bibnamefont {Gnan}},\ }\href {\doibase
  10.1039/D0SM02162H} {\bibfield  {journal} {\bibinfo  {journal} {Soft Matter}\
  }\textbf {\bibinfo {volume} {17}},\ \bibinfo {pages} {3807} (\bibinfo {year}
  {2021})}\BibitemShut {NoStop}%
\bibitem [{\citenamefont {Dittrich}\ \emph {et~al.}(2021)\citenamefont
  {Dittrich}, \citenamefont {Speck},\ and\ \citenamefont
  {Virnau}}]{Dittrich2021}%
  \BibitemOpen
  \bibfield  {author} {\bibinfo {author} {\bibfnamefont {F.}~\bibnamefont
  {Dittrich}}, \bibinfo {author} {\bibfnamefont {T.}~\bibnamefont {Speck}}, \
  and\ \bibinfo {author} {\bibfnamefont {P.}~\bibnamefont {Virnau}},\ }\href
  {\doibase 10.1140/epje/s10189-021-00058-1} {\bibfield  {journal} {\bibinfo
  {journal} {The European Physical Journal E}\ }\textbf {\bibinfo {volume}
  {44}},\ \bibinfo {pages} {53} (\bibinfo {year} {2021})}\BibitemShut {NoStop}%
\bibitem [{\citenamefont {Ray}\ \emph {et~al.}(2024)\citenamefont {Ray},
  \citenamefont {Mukherjee},\ and\ \citenamefont {Mohanty}}]{Ray_2024}%
  \BibitemOpen
  \bibfield  {author} {\bibinfo {author} {\bibfnamefont {C.~G.}\ \bibnamefont
  {Ray}}, \bibinfo {author} {\bibfnamefont {I.}~\bibnamefont {Mukherjee}}, \
  and\ \bibinfo {author} {\bibfnamefont {P.~K.}\ \bibnamefont {Mohanty}},\
  }\href {\doibase 10.1088/1742-5468/ad685b} {\bibfield  {journal} {\bibinfo
  {journal} {Journal of Statistical Mechanics: Theory and Experiment}\ }\textbf
  {\bibinfo {volume} {2024}},\ \bibinfo {pages} {093207} (\bibinfo {year}
  {2024})}\BibitemShut {NoStop}%
\bibitem [{\citenamefont {Fortunato}(2002)}]{PhysRevB.66.054107}%
  \BibitemOpen
  \bibfield  {author} {\bibinfo {author} {\bibfnamefont {S.}~\bibnamefont
  {Fortunato}},\ }\href {\doibase 10.1103/PhysRevB.66.054107} {\bibfield
  {journal} {\bibinfo  {journal} {Phys. Rev. B}\ }\textbf {\bibinfo {volume}
  {66}},\ \bibinfo {pages} {054107} (\bibinfo {year} {2002})}\BibitemShut
  {NoStop}%
\bibitem [{\citenamefont {Stella}\ and\ \citenamefont
  {Vanderzande}(1989)}]{PhysRevLett.62.1067}%
  \BibitemOpen
  \bibfield  {author} {\bibinfo {author} {\bibfnamefont {A.~L.}\ \bibnamefont
  {Stella}}\ and\ \bibinfo {author} {\bibfnamefont {C.}~\bibnamefont
  {Vanderzande}},\ }\href {\doibase 10.1103/PhysRevLett.62.1067} {\bibfield
  {journal} {\bibinfo  {journal} {Phys. Rev. Lett.}\ }\textbf {\bibinfo
  {volume} {62}},\ \bibinfo {pages} {1067} (\bibinfo {year}
  {1989})}\BibitemShut {NoStop}%
\bibitem [{\citenamefont {Bonati}\ \emph {et~al.}(2019)\citenamefont {Bonati},
  \citenamefont {Pelissetto},\ and\ \citenamefont
  {Vicari}}]{PhysRevLett.123.232002}%
  \BibitemOpen
  \bibfield  {author} {\bibinfo {author} {\bibfnamefont {C.}~\bibnamefont
  {Bonati}}, \bibinfo {author} {\bibfnamefont {A.}~\bibnamefont {Pelissetto}},
  \ and\ \bibinfo {author} {\bibfnamefont {E.}~\bibnamefont {Vicari}},\ }\href
  {\doibase 10.1103/PhysRevLett.123.232002} {\bibfield  {journal} {\bibinfo
  {journal} {Phys. Rev. Lett.}\ }\textbf {\bibinfo {volume} {123}},\ \bibinfo
  {pages} {232002} (\bibinfo {year} {2019})}\BibitemShut {NoStop}%
\bibitem [{\citenamefont {Mukherjee}\ and\ \citenamefont
  {Mohanty}(2023)}]{PhysRevB.108.174417}%
  \BibitemOpen
  \bibfield  {author} {\bibinfo {author} {\bibfnamefont {I.}~\bibnamefont
  {Mukherjee}}\ and\ \bibinfo {author} {\bibfnamefont {P.~K.}\ \bibnamefont
  {Mohanty}},\ }\href {\doibase 10.1103/PhysRevB.108.174417} {\bibfield
  {journal} {\bibinfo  {journal} {Phys. Rev. B}\ }\textbf {\bibinfo {volume}
  {108}},\ \bibinfo {pages} {174417} (\bibinfo {year} {2023})}\BibitemShut
  {NoStop}%
\bibitem [{\citenamefont {Paoluzzi}\ \emph {et~al.}(2020)\citenamefont
  {Paoluzzi}, \citenamefont {Maggi},\ and\ \citenamefont
  {Crisanti}}]{PhysRevResearch.2.023207}%
  \BibitemOpen
  \bibfield  {author} {\bibinfo {author} {\bibfnamefont {M.}~\bibnamefont
  {Paoluzzi}}, \bibinfo {author} {\bibfnamefont {C.}~\bibnamefont {Maggi}}, \
  and\ \bibinfo {author} {\bibfnamefont {A.}~\bibnamefont {Crisanti}},\ }\href
  {\doibase 10.1103/PhysRevResearch.2.023207} {\bibfield  {journal} {\bibinfo
  {journal} {Phys. Rev. Res.}\ }\textbf {\bibinfo {volume} {2}},\ \bibinfo
  {pages} {023207} (\bibinfo {year} {2020})}\BibitemShut {NoStop}%
\bibitem [{\citenamefont {Stauffer}(1979)}]{STAUFFER19791}%
  \BibitemOpen
  \bibfield  {author} {\bibinfo {author} {\bibfnamefont {D.}~\bibnamefont
  {Stauffer}},\ }\href {\doibase https://doi.org/10.1016/0370-1573(79)90060-7}
  {\bibfield  {journal} {\bibinfo  {journal} {Physics Reports}\ }\textbf
  {\bibinfo {volume} {54}},\ \bibinfo {pages} {1} (\bibinfo {year}
  {1979})}\BibitemShut {NoStop}%
\bibitem [{\citenamefont {Janke}\ and\ \citenamefont
  {Schakel}(2005)}]{PhysRevE.71.036703}%
  \BibitemOpen
  \bibfield  {author} {\bibinfo {author} {\bibfnamefont {W.}~\bibnamefont
  {Janke}}\ and\ \bibinfo {author} {\bibfnamefont {A.~M.~J.}\ \bibnamefont
  {Schakel}},\ }\href {\doibase 10.1103/PhysRevE.71.036703} {\bibfield
  {journal} {\bibinfo  {journal} {Phys. Rev. E}\ }\textbf {\bibinfo {volume}
  {71}},\ \bibinfo {pages} {036703} (\bibinfo {year} {2005})}\BibitemShut
  {NoStop}%
\bibitem [{\citenamefont {Essam}(1980)}]{Essam_1980}%
  \BibitemOpen
  \bibfield  {author} {\bibinfo {author} {\bibfnamefont {J.~W.}\ \bibnamefont
  {Essam}},\ }\href {\doibase 10.1088/0034-4885/43/7/001} {\bibfield  {journal}
  {\bibinfo  {journal} {Reports on Progress in Physics}\ }\textbf {\bibinfo
  {volume} {43}},\ \bibinfo {pages} {833} (\bibinfo {year} {1980})}\BibitemShut
  {NoStop}%
\bibitem [{\citenamefont {Margolina}\ \emph {et~al.}(1982)\citenamefont
  {Margolina}, \citenamefont {Herrmann},\ and\ \citenamefont
  {Stauffer}}]{MARGOLINA198273}%
  \BibitemOpen
  \bibfield  {author} {\bibinfo {author} {\bibfnamefont {A.}~\bibnamefont
  {Margolina}}, \bibinfo {author} {\bibfnamefont {H.}~\bibnamefont {Herrmann}},
  \ and\ \bibinfo {author} {\bibfnamefont {D.}~\bibnamefont {Stauffer}},\
  }\href {\doibase https://doi.org/10.1016/0375-9601(82)90219-5} {\bibfield
  {journal} {\bibinfo  {journal} {Physics Letters A}\ }\textbf {\bibinfo
  {volume} {93}},\ \bibinfo {pages} {73} (\bibinfo {year} {1982})}\BibitemShut
  {NoStop}%
\bibitem [{\citenamefont {Binder}\ and\ \citenamefont
  {Heermann}(2010)}]{binder2010monte}%
  \BibitemOpen
  \bibfield  {author} {\bibinfo {author} {\bibfnamefont {K.}~\bibnamefont
  {Binder}}\ and\ \bibinfo {author} {\bibfnamefont {D.~W.}\ \bibnamefont
  {Heermann}},\ }\href {\doibase 10.1007/978-3-642-03163-2} {\emph {\bibinfo
  {title} {Monte Carlo Simulation in Statistical Physics: An Introduction}}},\
  \bibinfo {edition} {5th}\ ed.,\ Graduate Texts in Physics\ (\bibinfo
  {publisher} {Springer Berlin, Heidelberg},\ \bibinfo {year} {2010})\ pp.\
  \bibinfo {pages} {XIV, 202},\ \bibinfo {note} {originally published as volume
  80 in the series: Springer Series in Solid-State Sciences, Published: 25
  August 2010 (Hardcover), 17 August 2010 (eBook)}\BibitemShut {NoStop}%
\bibitem [{\citenamefont {Binder}(1981)}]{PhysRevLett.47.693}%
  \BibitemOpen
  \bibfield  {author} {\bibinfo {author} {\bibfnamefont {K.}~\bibnamefont
  {Binder}},\ }\href {\doibase 10.1103/PhysRevLett.47.693} {\bibfield
  {journal} {\bibinfo  {journal} {Phys. Rev. Lett.}\ }\textbf {\bibinfo
  {volume} {47}},\ \bibinfo {pages} {693} (\bibinfo {year} {1981})}\BibitemShut
  {NoStop}%
\bibitem [{\citenamefont {Privman}(1990)}]{doi:10.1142/1011}%
  \BibitemOpen
  \bibfield  {author} {\bibinfo {author} {\bibfnamefont {V.}~\bibnamefont
  {Privman}},\ }\href {\doibase 10.1142/1011} {\emph {\bibinfo {title} {Finite
  Size Scaling and Numerical Simulation of Statistical Systems}}}\ (\bibinfo
  {publisher} {WORLD SCIENTIFIC},\ \bibinfo {year} {1990})\ \Eprint
  {http://arxiv.org/abs/https://www.worldscientific.com/doi/pdf/10.1142/1011}
  {https://www.worldscientific.com/doi/pdf/10.1142/1011} \BibitemShut {NoStop}%
\bibitem [{\citenamefont {Landau}\ and\ \citenamefont
  {Binder}(2014)}]{Landau_Binder_2014}%
  \BibitemOpen
  \bibfield  {author} {\bibinfo {author} {\bibfnamefont {D.~P.}\ \bibnamefont
  {Landau}}\ and\ \bibinfo {author} {\bibfnamefont {K.}~\bibnamefont
  {Binder}},\ }\href@noop {} {\emph {\bibinfo {title} {A Guide to Monte Carlo
  Simulations in Statistical Physics}}},\ \bibinfo {edition} {4th}\ ed.\
  (\bibinfo  {publisher} {Cambridge University Press},\ \bibinfo {year}
  {2014})\BibitemShut {NoStop}%
\bibitem [{\citenamefont {Luijten}\ \emph {et~al.}(2002)\citenamefont
  {Luijten}, \citenamefont {Fisher},\ and\ \citenamefont
  {Panagiotopoulos}}]{PhysRevLett.88.185701}%
  \BibitemOpen
  \bibfield  {author} {\bibinfo {author} {\bibfnamefont {E.}~\bibnamefont
  {Luijten}}, \bibinfo {author} {\bibfnamefont {M.~E.}\ \bibnamefont {Fisher}},
  \ and\ \bibinfo {author} {\bibfnamefont {A.~Z.}\ \bibnamefont
  {Panagiotopoulos}},\ }\href {\doibase 10.1103/PhysRevLett.88.185701}
  {\bibfield  {journal} {\bibinfo  {journal} {Phys. Rev. Lett.}\ }\textbf
  {\bibinfo {volume} {88}},\ \bibinfo {pages} {185701} (\bibinfo {year}
  {2002})}\BibitemShut {NoStop}%
\bibitem [{\citenamefont {Vanderzande}(1992)}]{Vanderzande_1992}%
  \BibitemOpen
  \bibfield  {author} {\bibinfo {author} {\bibfnamefont {C.}~\bibnamefont
  {Vanderzande}},\ }\href {\doibase 10.1088/0305-4470/25/2/008} {\bibfield
  {journal} {\bibinfo  {journal} {Journal of Physics A: Mathematical and
  General}\ }\textbf {\bibinfo {volume} {25}},\ \bibinfo {pages} {L75}
  (\bibinfo {year} {1992})}\BibitemShut {NoStop}%
\bibitem [{\citenamefont {Marro}\ \emph {et~al.}(1987)\citenamefont {Marro},
  \citenamefont {Vall\'es},\ and\ \citenamefont
  {Gonz\'alez-Miranda}}]{PhysRevB.35.3372}%
  \BibitemOpen
  \bibfield  {author} {\bibinfo {author} {\bibfnamefont {J.}~\bibnamefont
  {Marro}}, \bibinfo {author} {\bibfnamefont {J.~L.}\ \bibnamefont {Vall\'es}},
  \ and\ \bibinfo {author} {\bibfnamefont {J.~M.}\ \bibnamefont
  {Gonz\'alez-Miranda}},\ }\href {\doibase 10.1103/PhysRevB.35.3372} {\bibfield
   {journal} {\bibinfo  {journal} {Phys. Rev. B}\ }\textbf {\bibinfo {volume}
  {35}},\ \bibinfo {pages} {3372} (\bibinfo {year} {1987})}\BibitemShut
  {NoStop}%
\bibitem [{\citenamefont {Albano}\ and\ \citenamefont
  {Saracco}(2002)}]{PhysRevLett.88.145701}%
  \BibitemOpen
  \bibfield  {author} {\bibinfo {author} {\bibfnamefont {E.~V.}\ \bibnamefont
  {Albano}}\ and\ \bibinfo {author} {\bibfnamefont {G.}~\bibnamefont
  {Saracco}},\ }\href {\doibase 10.1103/PhysRevLett.88.145701} {\bibfield
  {journal} {\bibinfo  {journal} {Phys. Rev. Lett.}\ }\textbf {\bibinfo
  {volume} {88}},\ \bibinfo {pages} {145701} (\bibinfo {year}
  {2002})}\BibitemShut {NoStop}%
\end{thebibliography}%

\end{document}


\title{Supplemental Material: Site-percolation transition  of run-and-tumble  particles}

\author{
Soumya K. Saha, 
Aikya Banerjee, and P. K. Mohanty}
\email{pkmohanty@iiserkol.ac.in}
\affiliation {Department of Physical Sciences, Indian Institute of Science Education and Research Kolkata, Mohanpur, 741246 India.}
\begin{abstract} 
The Supplemental Material  provides additional results to  support and  strengthen  the  claims  of the article.  First we  discuss  the simulation methods  and  describe  the evolution of the RTP system  to  the   steady state. We also provide  the   finite size scaling collapse  for  all the  critical points, mentioned in the main text.  
\end{abstract} 
\maketitle

\section{Steady state  of the model}

\begin{figure}[h]
    \centering
    \begin{subfigure}[b]{0.325\textwidth}
        \centering
        \includegraphics[width=\textwidth]{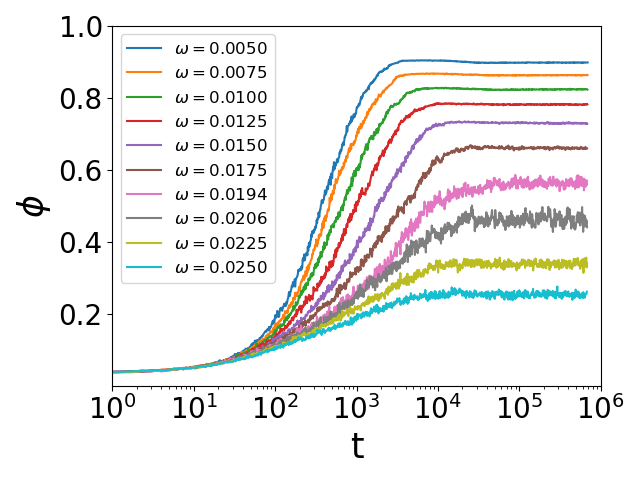}
        \caption{}
        \label{fig:figure1}
    \end{subfigure}
    \hfill
    \begin{subfigure}[b]{0.325\textwidth}
        \centering
        \includegraphics[width=\textwidth]{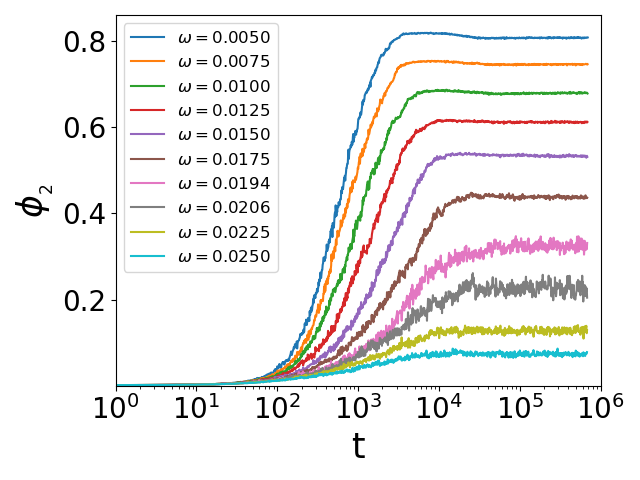}
        \caption{}
        \label{fig:figure2}
    \end{subfigure}
    \hfill
    \begin{subfigure}[b]{0.325\textwidth}
        \centering
        \includegraphics[width=\textwidth]{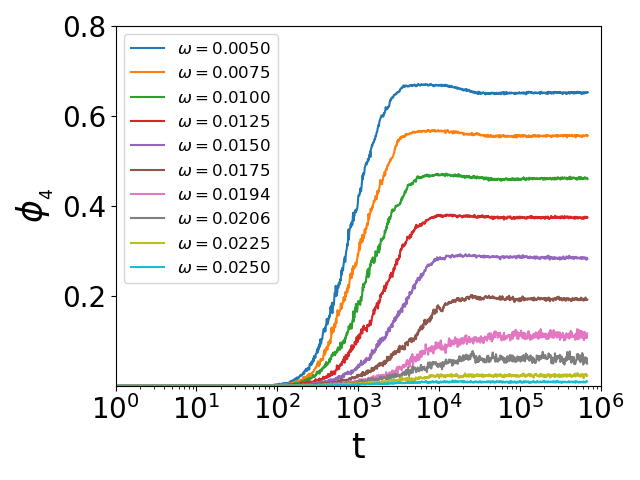}
        \caption{}
        \label{fig:figure3}
    \end{subfigure}
    \caption{Time evolution of (a)$\phi$, (b)$\phi_2$, and  (c)$\phi_4$ for $p=0.235$ 
 and various values of $\omega$, starting from a random initial distribution of $N=L^2/2$ RTPs.  Here $\phi_k$ is the $k$-th moment of the largest cluster and is given by $\phi_k= \frac{1} {N^k} \langle s_{max}^k\rangle.$  Here $L=128$ and the data  is averaged over $200$ initial conditions.  }
    \label{fig:F10}
\end{figure}

From the Monte Carlo simulations of the model defined in Eq. (2)  in main text,   we  calculate   the steady state averages of the largest cluster 
$\phi=\frac{1} {N} \langle s_{max}\rangle$ and the $k$-th  moment, 
$\phi_k= \frac{1} {N^k} \langle s_{max}^k\rangle.$  A cluster is defined   in a way similar to the clusters  defined in the site percolation problem - two particles separated by one lattice  unit  belong to  same cluster. 
In  Fig. \ref{fig:F10}  we show   for a fixed $p=0.235$ how  $\phi, \phi_2$ and $\phi_4$ evolve   with  time $t$  with different values of $\omega,$ starting from a random initial distribution of RTPs    on a  $L\times L$ square lattice  ($L=128$).  The initial orientation of particles are also chosen  randomly and independently.  It  turns out that   $s_{max}$ and  its  moments reach their stationary value   well  before $t=10^5.$   This gives us  an estimate of  the relaxation time. In all  the simulations we  relax the systems   until $t=10^5$  and then  calculate the steady state average values 
in  next  $10^6$  MCS.  The steady state  data is  further averaged over  $200$  different initial conditions.  From these steady state 
 values   we  obtain   the order parameter  $\phi$,  the susceptibility $\chi = \phi_2 -\phi^2$ and  the Binder cumulant $U_4  =  1- \frac{\phi_4} {3\phi_2^2}.$

\subsection{The phase diagram}
\begin{figure}[h]
    \centering
    \includegraphics[width=.8\linewidth]{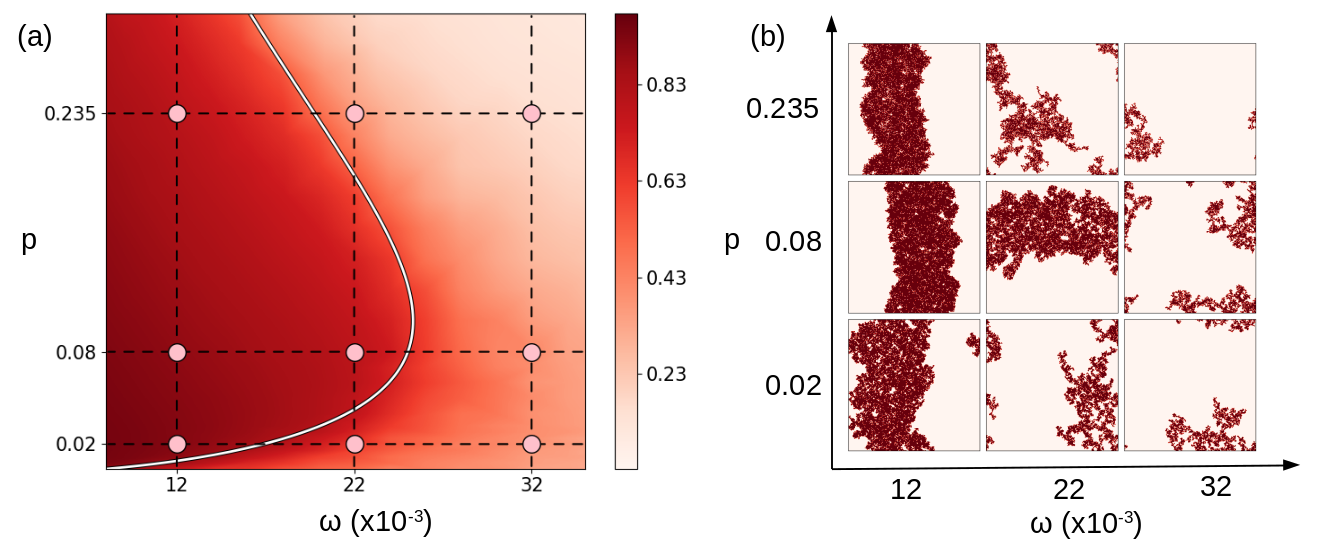}
    \caption{(a) Phase diagram:  The  solid line, obtained from the best fit of the critical points,  separates the  mixed phase from phase separated state.  The background   is  the heat  map of  the order parameter.   (b)  Typical largest clusters  at  nine different    points  $(p=0.02,0.08, 0.235)$ $\times$   $(\omega = 0.012, 0.022, 0.032)$   marked as circles in (a). }
    \label{fig:enter-label}
\end{figure}

To obtain  the  phase diagram   we  plot   the   heat map  of the order parameter  $\langle s_{max} \rangle $ in $(\omega,p)$- plane  in Fig. \ref{fig:enter-label}(a) where the darker region  represent a phase separated state  that contains  a macroscopic-cluster. The line  that separates the    mixed phase from the phase-separated state is obtained   from  a best fit of the the critical points obtained from simulations.  Clearly,  a re-entrant  percolation transition is observed  when  
$p$  is varied  keeping $\omega$ fixed.   With increased  $p$   percolated state appears  and  then  then it disappears   with   further increase of $p$.   Typical largest clusters of the system  are shown  in  Fig. \ref{fig:enter-label}(b)
at nine different points $(p=0.02,0.08, 0.235)$ $\times$   $(\omega = 0.012, 0.022, 0.032)$   marked as circles.   The  largest cluster  is macroscopic (and spans the lattice)  for all three values of $p$  when $\omega=0.012.$ Whereas for  $\omega=0.022,$ the largest cluster is  small  for  $p=0.08,$   it becomes much denser  at   $p=0.022$  and  becomes  sparsely connected  again   at  a  larger value of $p=0.235$  indicating  the  re-entrant nature of the transition. 

\subsection{Critical exponents from  finite size scaling}

 \begin{figure}[h]
\centering
\includegraphics[width=5.65cm]{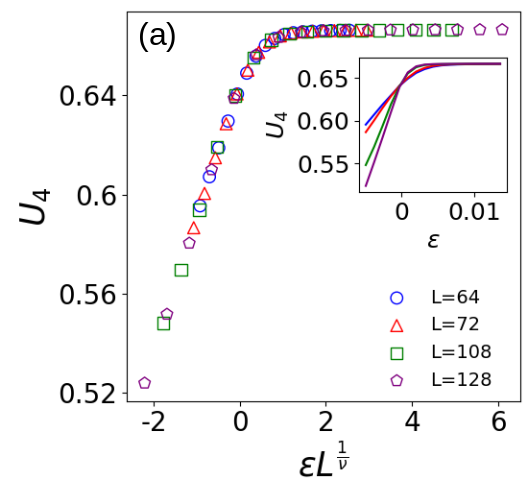}
\includegraphics[width=5.65cm]{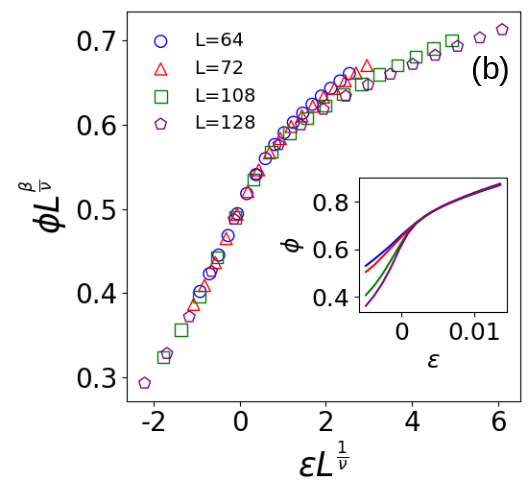}
\includegraphics[width=5.65cm]{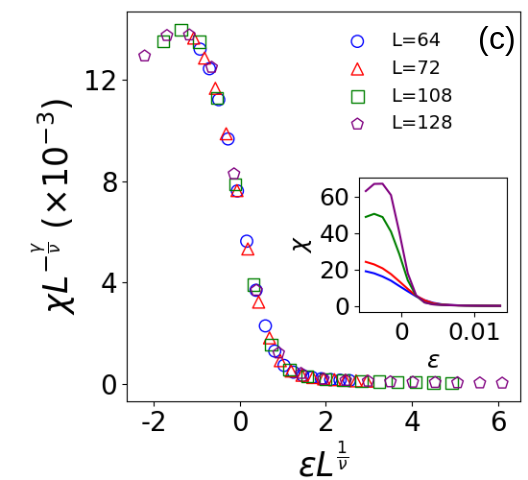}
\caption{Critical point II. $(p_c,\omega_c)=(0.150,0.0235):$  (a)$U_4$, (b) $\phi L^\frac\beta\nu$ and (c)$\chi L^{-\frac{\gamma}{\nu}}$ as a function of  $\varepsilon L^\frac{1}{\nu}$. The  best collapse  is obtained for  $\frac1\nu=1.26,$ $\frac\beta\nu=0.10,$  $\frac\gamma\nu=1.75.$  The inset shows  raw data, $U_4, \phi,\chi$ vs. $\varepsilon.$}
\label{fig:p0.15}
\end{figure}

\begin{figure}[h]
\centering
\includegraphics[width=5.65cm]{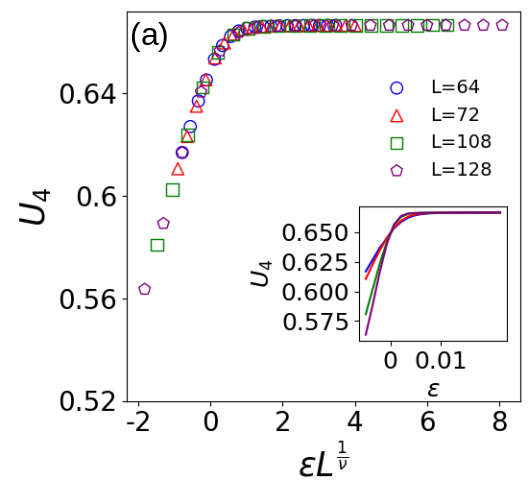}
\includegraphics[width=5.65cm]{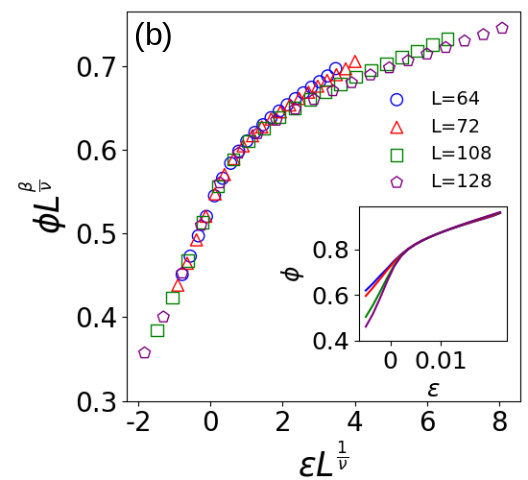}
\includegraphics[width=5.65cm]{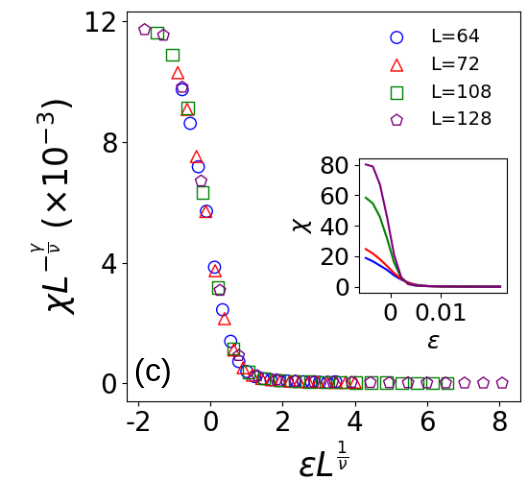}
\caption{Critical point III. $(p_c,\omega_c)=(0.080,0.0247):$  (a)$U_4$, (b) $\phi L^\frac\beta\nu$ and (c)$\chi L^{-\frac{\gamma}{\nu}}$ as a function of  $\varepsilon L^\frac{1}{\nu}$. The  best collapse  is obtained for  $\frac1\nu=1.22,$ $\frac\beta\nu=0.09,$  $\frac\gamma\nu=1.82.$  The inset shows  raw data, $U_4, \phi,\chi$ vs. $\varepsilon.$}
\label{fig:p0.08}
\end{figure}
\begin{figure}[h]
\centering
\includegraphics[width=5.65cm]{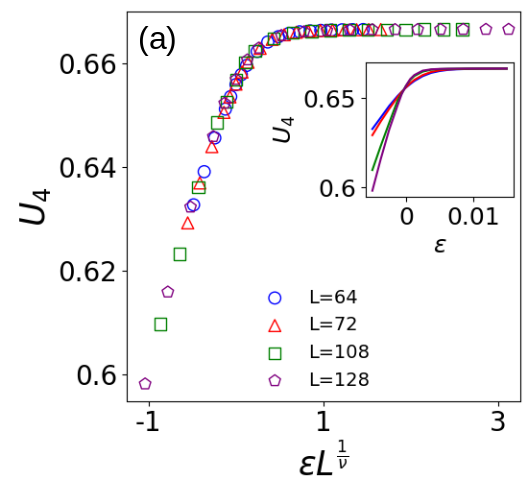}
\includegraphics[width=5.65cm]{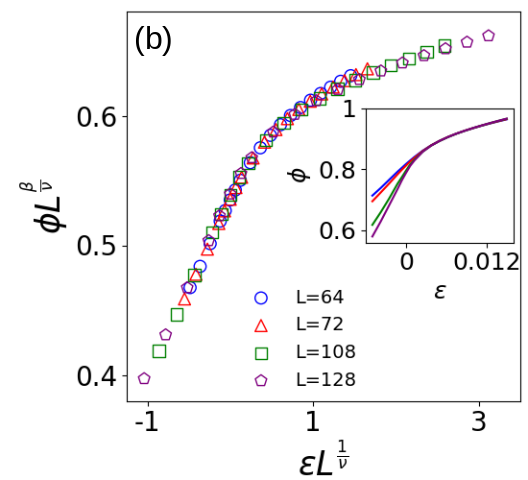}
\includegraphics[width=5.65cm]{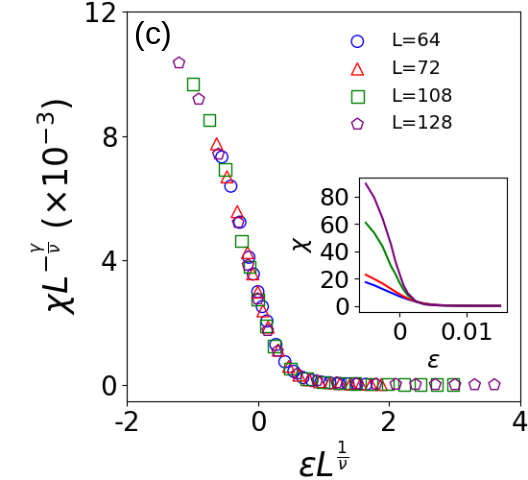}
\caption{Critical point IV. $(p_c,\omega_c)=(0.029,0.020):$  (a)$U_4$, (b) $\phi L^\frac\beta\nu$ and (c)$\chi L^{-\frac{\gamma}{\nu}}$ as a function of  $\varepsilon L^\frac{1}{\nu}$. The  best collapse  is obtained for  $\frac1\nu=1.13,$ $\frac\beta\nu=0.066,$  $\frac\gamma\nu=1.868.$  The inset shows  raw data, $U_4, \phi,\chi$ vs. $\varepsilon.$}
\label{fig:p0.029}
\end{figure}
\begin{figure}[h]
\centering
\includegraphics[width=5.65cm]{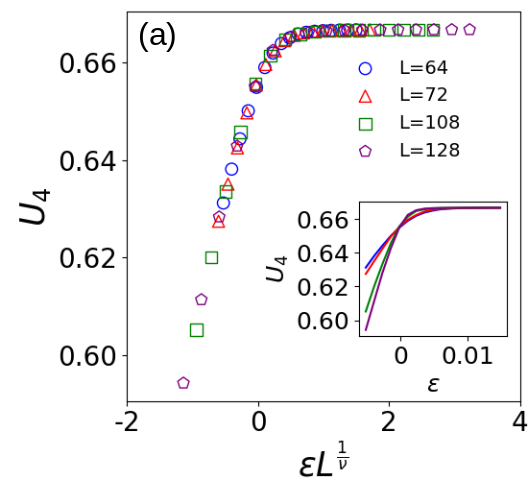}
\includegraphics[width=5.65cm]{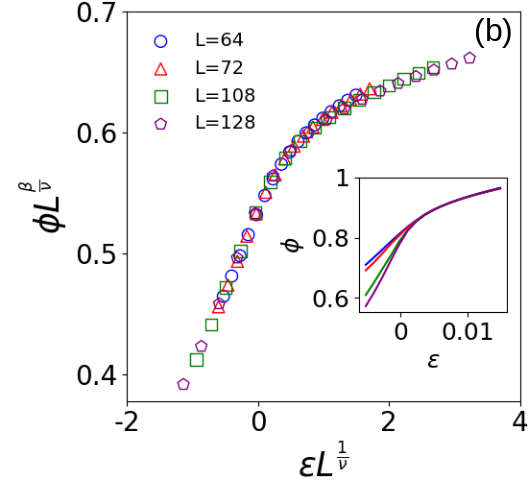}
\includegraphics[width=5.65cm]{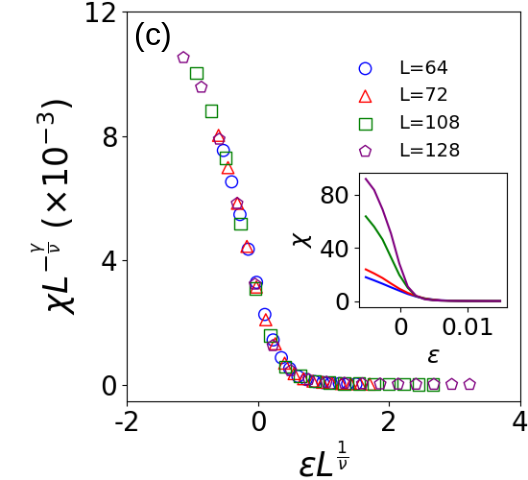}
\caption{Critical point V. $(p_c,\omega_c)=(0.0275,0.019):$  (a)$U_4$, (b) $\phi L^\frac\beta\nu$ and (c)$\chi L^{-\frac{\gamma}{\nu}}$ as a function of  $\varepsilon L^\frac{1}{\nu}$. The  best collapse  is obtained for  $\frac1\nu=1.11,$ $\frac\beta\nu=0.065,$  $\frac\gamma\nu=1.87.$  The inset shows  raw data, $U_4, \phi,\chi$ vs. $\varepsilon.$}
\label{fig:p0.0275}
\end{figure}
\begin{figure}[h]
\centering
\includegraphics[width=5.65cm]{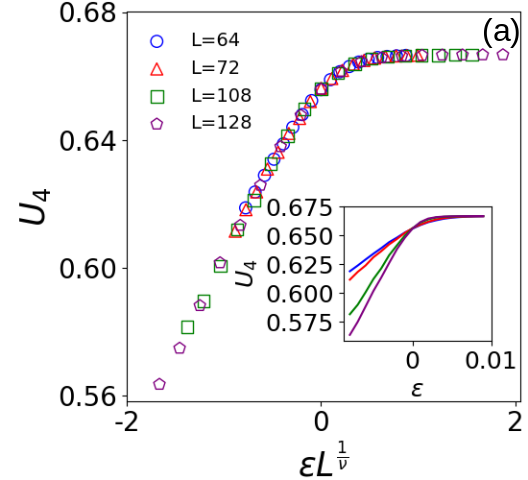}
\includegraphics[width=5.65cm]{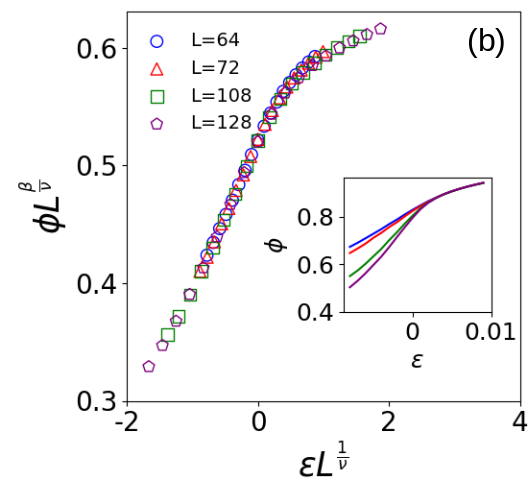}
\includegraphics[width=5.65cm]{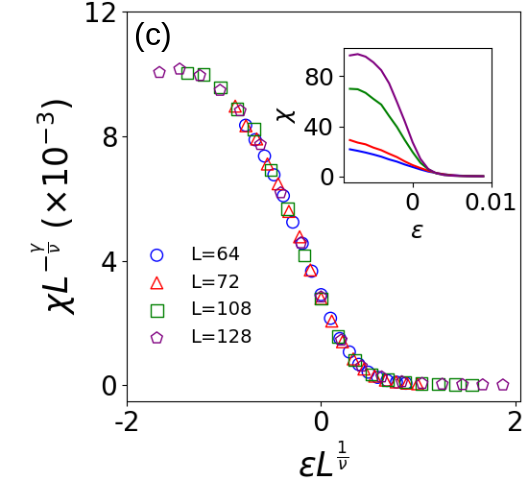}
\caption{Critical point VI. $(p_c,\omega_c)=(0.020,0.018):$  (a)$U_4$, (b) $\phi L^\frac\beta\nu$ and (c)$\chi L^{-\frac{\gamma}{\nu}}$ as a function of  $\varepsilon L^\frac{1}{\nu}$. The  best collapse  is obtained for  $\frac1\nu=1.10,$ $\frac\beta\nu=0.055,$  $\frac\gamma\nu=1.89.$  The inset shows  raw data, $U_4, \phi,\chi$ vs. $\varepsilon.$}
\label{fig:p0.02}
\end{figure}

To obtain the static critical exponents  $\nu,\beta,\gamma$ we   employ  the 
 finite size scaling analysis,
 \begin{eqnarray}\label{eq:scaling_binder}
 \phi = L^{-\frac{\beta}{\nu}}f_\phi(\varepsilon L^{\frac1\nu}); ~ 
 \chi = L^{\frac\gamma\nu}f_\chi(\varepsilon L^{\frac1\nu});~
  U_4 = f_{b}(\varepsilon L^{\frac1\nu}).
\end{eqnarray}
For  one of the critical point I. $(p_c,\omega_c)=(0.235,0.020)$   
Figs. 3(a), (b) and (c) in the main text show  respectively the scaling collapse of $\phi  L^{\frac{\beta}{\nu}}$, $\chi  L^{-\frac\gamma\nu}$ and  $U_4$ as a function of $\varepsilon L^{\frac1\nu}.$   Similar scaling collapse  for the other critical points  (II to VI mentioned in Table I of the  main text) are shown respectively in Figs. \ref{fig:p0.15}  to \ref{fig:p0.02}. 

\subsection{Critical behaviour  in  $p$- direction}
\begin{figure}[h]
\centering
\includegraphics[width=5.65cm]{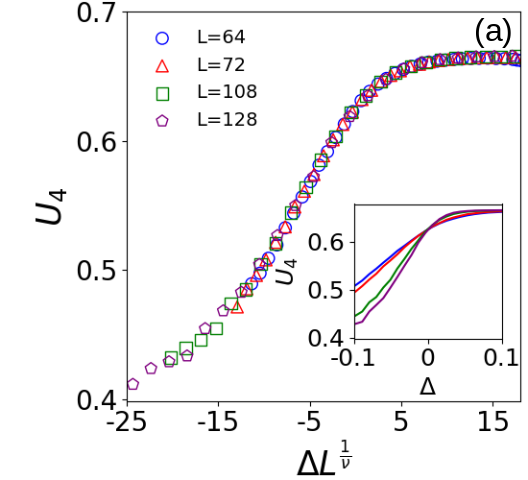}
\includegraphics[width=5.65cm]{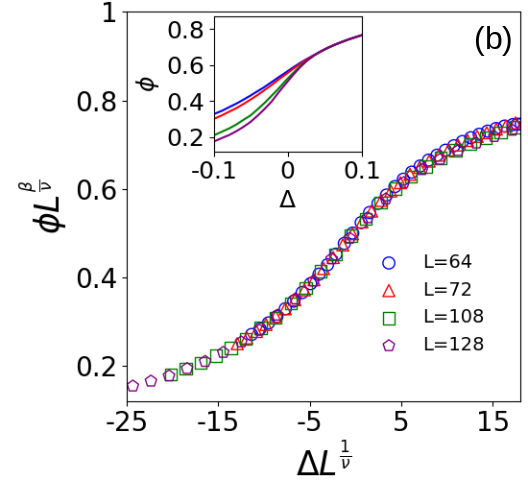}
\includegraphics[width=5.65cm]{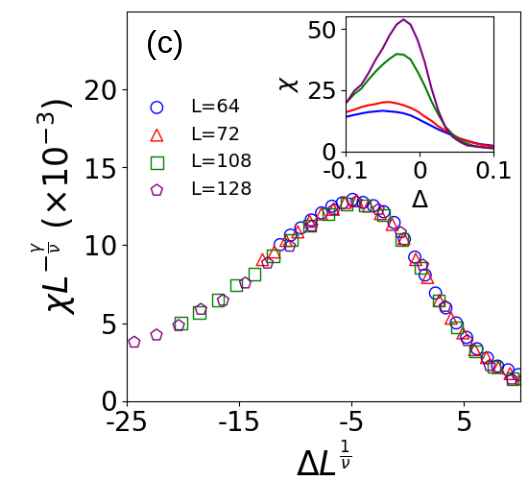}
\caption{Critical point IV. $(p_c,\omega_c)=(0.235,0.020):$  (a)$U_4$, (b) $\phi L^\frac\beta\nu$ and (c)$\chi L^{-\frac{\gamma}{\nu}}$ as a function of  $\Delta L^\frac{1}{\nu}$. The  best collapse  is obtained for  $\frac1\nu=1.43,$ $\frac\beta\nu=0.140,$  $\frac\gamma\nu=1.72.$  The inset shows  raw data, $U_4, \phi,\chi$ vs. $\Delta.$}
\label{fig:w0.235}
\end{figure}

In a  two parameter space there are  two independent direction. At any critical point the  critical  exponents  can  be obtained  from varying one of the paramters keeping the other fixed. So far we  have  done the finite size scaling analysis   by varying $\omega$  for a fixed  $p$. For consistency, we  consider  one of the  critical point   $(p_c,\omega_c)=(0.235,0.020),$   fix   $\omega=0.02$  and   study the critical behaviour by varying $p.$  Now the scaling functions depend on $\Delta = p_c-p,$  
 \begin{eqnarray}\label{eq:scaling_binder}
 \phi = L^{-\frac{\beta}{\nu}}\tilde f_\phi(\Delta L^{\frac1\nu}); ~ 
 \chi = L^{\frac\gamma\nu}\tilde f_\chi(\Delta L^{\frac1\nu});~
  U_4 = \tilde f_{b}(\Delta L^{\frac1\nu}).
\end{eqnarray}
A plot   of   $\phi  L^{\frac{\beta}{\nu}}$, $\chi  L^{-\frac\gamma\nu}$ and  $U_4$
as a function of $\Delta L^{\frac1\nu}$   is shown  in Fig. \ref{fig:w0.235} where 
$\frac{\beta}{\nu}$, $\frac\gamma\nu$  and  $\frac1\nu$  are   tuned  to a value that   gives  the  best collapse. We find that  the  critical  exponents   $\frac{1}{\nu}=1.43,\frac{\beta}{\nu}=0.14,\frac{\gamma}{\nu}=1.72$ obtained  previously by varying $\varepsilon,$   gives rise to  best data collapse.

\section{Motility Induced Phase separation transition }

\begin{figure}[h]
    \centering
    \includegraphics[width=.8\linewidth]{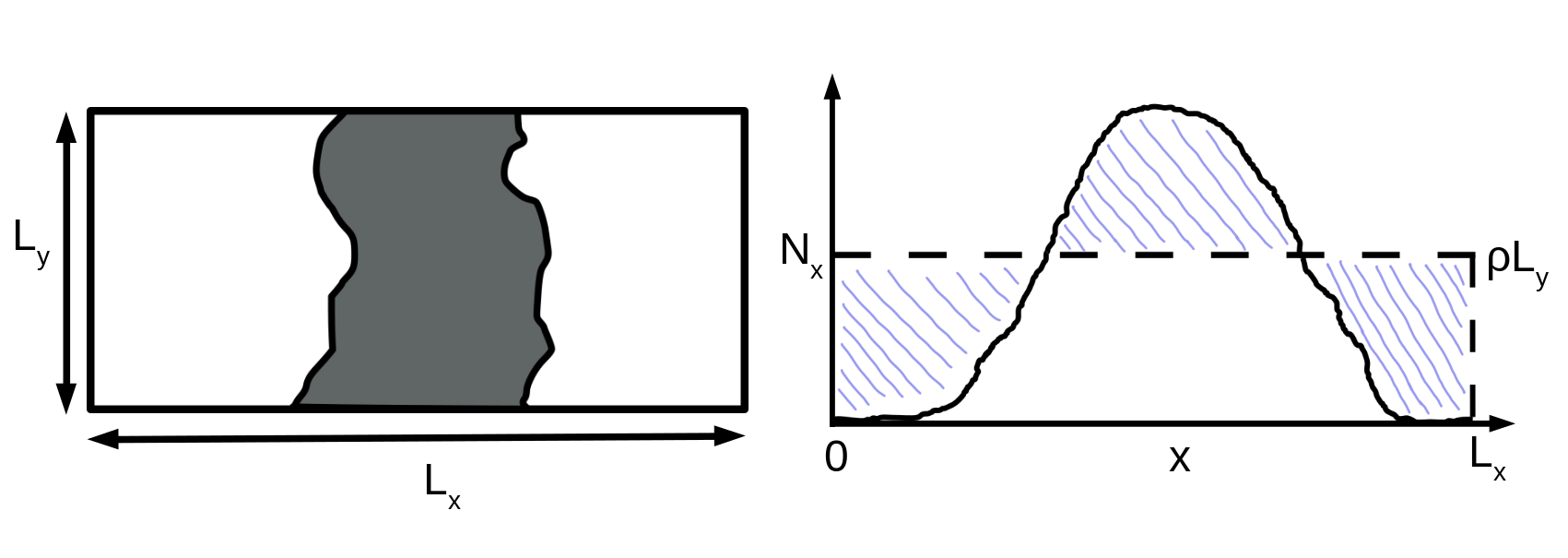}
    \caption{Schematic configuration of a phase-separated
state on a rectangular lattice ($L_x = 2L_y$ ).  (b) The order parameter $\phi'$
of the system measures how different is $N_x$ from its mean $\rho L_y$
in an absolute sense (the shaded area). Here $N_x$ counts the
total number of particles at all the lattice sites ${\bf i }\equiv (x, y)$ which have same $x$-coordinate.}
    \label{fig:sop}
\end{figure}

The RTPs  undergo a  percolation transition  when  $\omega$  is lowered below  a critical  threshold  value  $\omega_c$   that depends on  $p.$  This percolation transition belong  to the super universality class of $Z_2$ percolation.  Should we expect that   the motility  induced  phase separation transition of RTPs also occur   occurs at $\omega_c$?  In fact,  in  2D Ising Model  a percolation transition occurs  exactly at the  same critical temperature $T_c$  where   the  system 
phase transit from   being  a para-magnet to a ferromagnet.  The critical behaviour of   percolation transition  form a new universality class called $Z_2$-percolation   which is different from Ising universality class (IUC).  In a similar way,  if MIPS occurs  in   RTP  model   when   $\omega$  is lowered below $\omega_c,$  what    could   be a  suitable  order parameter to characterize such a transition? We use the order parameter suggested  in Ref.\cite{Ray}.
It is well-known that  in  a rectangular system   of  length  $L_x$ and height $L_y,$  the high density  phase boundaries     align in the shorter   direction.   Fig.  \ref{fig:sop}(a) shows  a typical  high density phase (shaded)    in a system where  $L_x= 2L_y.$   Then  the   order parameter $\phi'$  is  defined  as 
  \cite{Ray},
\be 
 \phi' = \frac{2}{L_x L_y} \sum_{x=1}^{L_x} \left| N_{x} - \rho L_y\right|;~ N_{x}= \sum_{y=1}^{L_y} n_{x,y},
\label{eq:OP}
\ee
where $N_x$  is  the total number of particles at lattice sites ${\bf i}\equiv (x,y)$ with the same $x$-coordinate.  A schematic representation of how  $\phi'$ quantifies a typical clustered configuration is shown in Fig \ref{fig:sop}.  

Using finite size scaling of the standard order parameter and the corresponding susceptibility at the critical points IV and VI (refer Table 1 of the main text) yields the exponents $\frac{\beta'}{\nu}$ and $\frac{\gamma'}{\nu}$  at those points. The relations between these exponents and the corresponding site-percolation exponents are then verified using Eq. (1) of the main text.